\documentclass[]{pasj01}

\Received{2020/10/21}
\Accepted{2021/01/08}
 
\usepackage{lscape}
\usepackage{color}
 
\begin{document} 

\title{ 
Spectrometer Using superconductor MIxer Receiver (SUMIRE) for Laboratory Submillimeter Spectroscopy}

\author{Yoshimasa \textsc{Watanabe}\altaffilmark{1,2,3,4}}%
\altaffiltext{1}{Materials Science and Engineering, College of Engineering, Shibaura Institute of Technology, 3-7-5 Toyosu, Koto-ku, Tokyo 135-8548, Japan}
\altaffiltext{2}{RIKEN Cluster for Pioneering Research, 2-1, Hirosawa, Wako, Saitama 351-0198, Japan}
\altaffiltext{3}{Division of Physics, Faculty of Pure and Applied Sciences, University of Tsukuba,  Tsukuba, Ibaraki 305-8571, Japan}
\altaffiltext{4}{College of Engineering, Nihon University, 1 Nakagawara, Tokusada, Tamuramachi, Koriyama, Fukushima 963-8642, Japan}

\author{Yutaro \textsc{Chiba},\altaffilmark{5,2}}
\altaffiltext{5}{Department of Physics, The University of Tokyo, 7-3-1 Hongo, Bunkyo-ku, Tokyo, 113-0033, Japan}

\author{Takeshi \textsc{Sakai}\altaffilmark{6}}
\altaffiltext{6}{Graduate School of Informatics and Engineering, The University of Electro-Communications, Chofu, Tokyo 182-8585, Japan}

\author{Akemi~\textsc{Tamanai}\altaffilmark{2}}

\author{Rikako \textsc{Suzuki}\altaffilmark{7}}
\altaffiltext{7}{Graduate School of Pure and Applied Sciences, University of Tsukuba,  Tsukuba, Ibaraki 305-8571, Japan}



\author{Nami \textsc{Sakai}\altaffilmark{2}}


\maketitle

\begin{abstract}
Recent spectroscopic observations by sensitive radio telescopes require accurate molecular spectral line frequencies to identify molecular species in a forest of lines detected.  To measure rest frequencies of molecular spectral lines in the laboratory, an emission-type millimeter and submillimeter-wave spectrometer utilizing state-of-the-art radio-astronomical technologies is developed.  The spectrometer is equipped with a 200~cm glass cylinder cell, a two sideband (2SB) Superconductor-Insulator-Superconductor (SIS) receiver in the 230~GHz band, and wide-band auto-correlation digital spectrometers.  By using the four 2.5~GHz digital spectrometers, a total instantaneous bandwidth of the 2SB SIS receiver of 8~GHz can be covered with a frequency resolution of 88.5~kHz.  Spectroscopic measurements of CH$_3$CN and HDO are carried out in the 230~GHz band so as to examine frequency accuracy, stability, sensitivity, as well as intensity calibration accuracy of our system.  As for the result of CH$_3$CN, we confirm that the frequency accuracy for lines detected with sufficient signal to noise ratio is better than 1~kHz, when the high resolution spectrometer having a channel resolution of 17.7~kHz is used.  In addition, we demonstrate the capability of this system by spectral scan measurement of CH$_3$OH from 216~GHz to 264~GHz.  We assign 242 transitions of CH$_3$OH, 51 transitions of $^{13}$CH$_3$OH, and 21 unidentified emission lines for 295 detected lines.  Consequently, our spectrometer demonstrates sufficient sensitivity, spectral resolution, and frequency accuracy for in-situ experimental-based rest frequency measurements of spectral lines on various molecular species. 

\end{abstract}

\section{Introduction}
More than 200 molecular species are now known as interstellar molecules, as complied in the Cologne Database for Molecular Spectroscopy (CDMS: \cite{muller01,cdms2005}), most of which are detected through radio-astronomical observations of rotational transition lines. Thanks to continuous development of sensitive receivers as well as wideband and high frequency-resolution spectrometers for radio astronomy, we can now scan a wide range of frequency and observe many spectral lines of various molecular species with reasonable observation time.  Taking advantage of this progress, many spectral line surveys have been conducted during the last two decades toward various types of astronomical sources such as hot cores (e.g., \cite{Schilke1997,Schilke2001,Tercero2010,Watanabe2015}), low-mass protostellar cores (e.g., \cite{Caux2011,Watanabe2012,Lindberg2015,Jorgensen2016,Yoshida2019}), dark clouds (e.g., \cite{Kaifu2004}), shocked regions (e.g., \cite{Sugimura2011,Yamaguchi2012}), Asymptotic Giant Branch stars (e.g., \cite{Cernicharo2000}), nearby galaxies (e.g.,\cite{Martin2006,Aladro2015,Takano2019,Watanabe2014,Watanabe2019}), and low-metallicity dwarf galaxies (e.g., \cite{Nishimura2016a,Nishimura2016b}).  These studies clearly reveal characteristic chemical composition of each source in an unbiased way, which allows us to discuss chemical and physical processes occurring there in detail. 

In high-angular-resolution and high-sensitivity spectroscopic observations with ALMA (Atacama Large Millimeter/submillimeter Array), various molecular species can readily be detected even in a single observation run. In these observations, many unidentified lines are also found (e.g., \cite{Imai2016,Watanabe2017,Belloche2019}).  Although most of them are likely high excitation lines and/or isotopologue lines of known interstellar molecules not listed in the molecular spectral line databases such as CDMS and Microwave Spectral Line Catalog provided by Jet Propulsion Laboratory (JPL) \citep{pickett98}, some of them could be spectral lines of new interstellar molecules.  Elimination of the weed lines of known molecules is essential for definitive identification of new species, and for this purpose, further enhancement of the spectral line database is awaited.  Besides, rest frequencies listed in the literature and/or the spectral line databases are not always as accurate as required for astronomical studies such as assignment of each line in a congested spectrum and kinematic analyses of a target astronomical source using the Doppler effect. For instance, the kinematic studies on disk formation around a newly formed protostar (e.g., \cite{Sakai2014,Oya2018,Watanabe2017}) require the accuracy better than 0.1~km~s$^{-1}$ (80~kHz at 230~GHz), which corresponds to the thermal linewidth at cold environments.  These problems are particularly serious for high excitation lines and isotopologue lines (e.g., \cite{Sakai2010,Sakai2013}).

Frequencies listed in the spectral line database are usually calculated by using molecular constants determined from the laboratory spectroscopic measurements in a limited range of rotational quantum numbers and a limited range of frequencies. Since the Hamiltonian for molecular rotation is represented by a kind of a power series of angular momentum operators, higher order terms are arbitrarily truncated in actual fitting of the measured frequencies due to vibration-rotation interactions\citep{Gordy1984}.  For this reason, the predicted frequencies of the transitions out of the observed ranges often suffer from large uncertainty and systematic deviations. This situation is particularly serious for spectral lines in the high frequency bands above 400 GHz (i.e., Band 8, Band 9, and Band 10 for ALMA), if their frequencies are based on extrapolation of the lower frequency measurements. For a full use of the data obtained by radio-astronomical observations, it is thus very useful to list up transition frequencies directly measured in the laboratory to complement the spectral line databases.

Based on this motivation, we have developed an emission-type millimeter and submillimeter spectrometer by introducing state-of-the-art technologies of radio astronomy. In contrast to ordinary absorption spectrometers (e.g., \cite{Petkie1997,Yamamoto1988}) and pulsed Fourier transform spectrometers (e.g., \cite{Ekkers1976,Balle1981,Brown2008}), this spectrometer observes thermal emission of molecules by using a superconducting receiver and wide band digital spectrometers. For the frontend, we employ the ALMA-type cartridge receiver, which is a two sideband (2SB) heterodyne receiver using two superconductor-insulator-superconductor (SIS) mixers.  A spectrometer of this type has an advantage in measuring transition frequencies accurately without modulation broadening as well as in measuring absolute intensities based on calibration using blackbody emission. Recently similar emission-type spectrometers have been developed in the millimeter and submillimeter bands with the same astronomical motivation mentioned above: the emission-type spectrometer for the 30-50 GHz and 71-115 GHz bands was developed by introducing high electron mobility transistor (HEMT) receivers for the Yebes 40~m radio telescope \citep{Tanarro2018,Cernicharo2019}, while that for the 270-390 GHz band was developed by using the SIS receiver at the University of Cologne \citep{Wehres2018}.  In this paper, we report the apparatus and the basic performance of our emission-type spectrometer covering the 210-270 GHz band and some demonstrative measurements for CH$_3$CN, HDO, and CH$_3$OH. 

\begin{figure*}[t]
 \begin{center}
  \includegraphics[width=15cm]{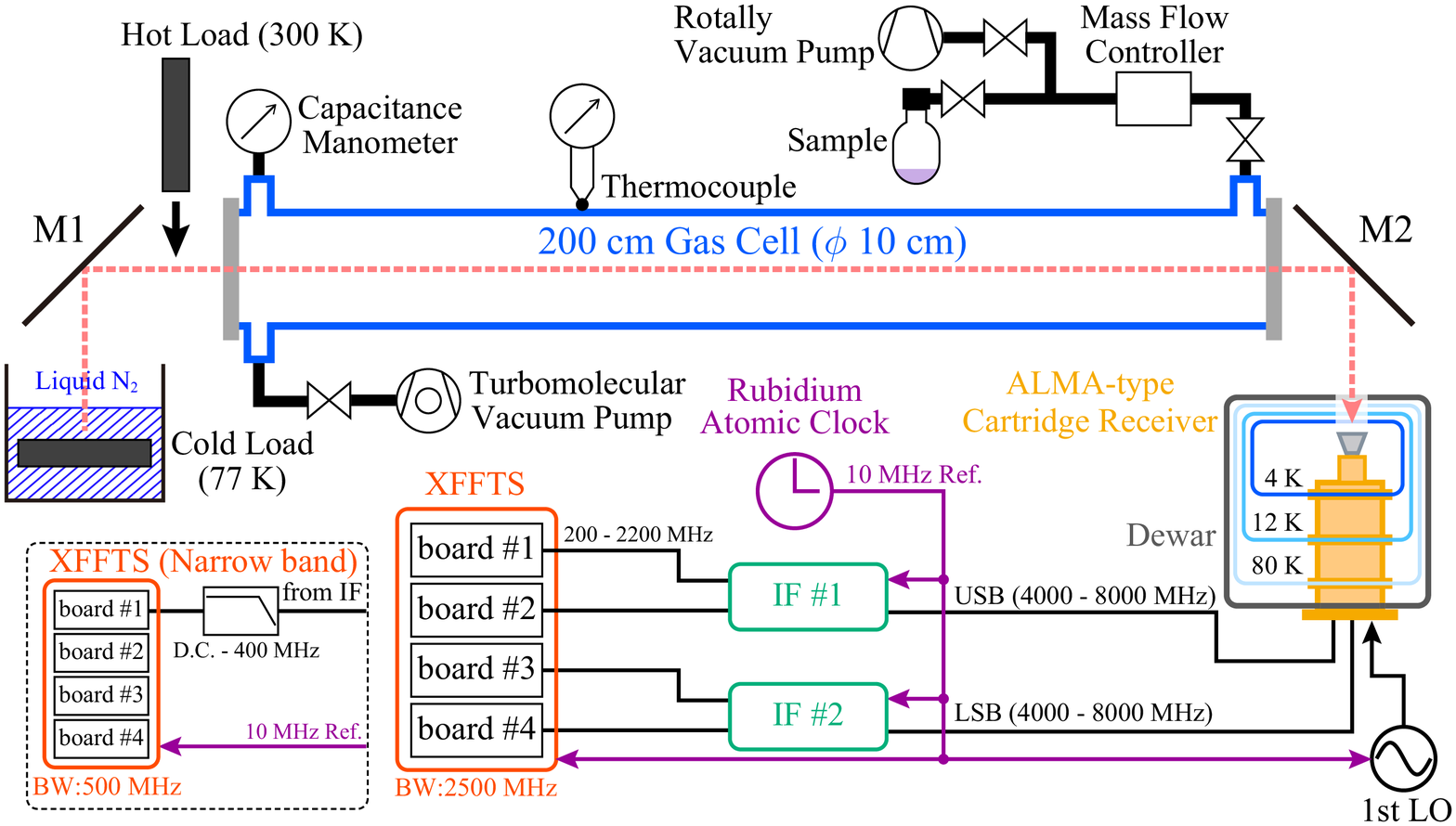}
 \end{center}
 \caption{Block diagram of the SUMIRE system.}
 \label{fig01}
\end{figure*}

\begin{figure*}[t]
 \begin{center}
  \includegraphics[width=15cm]{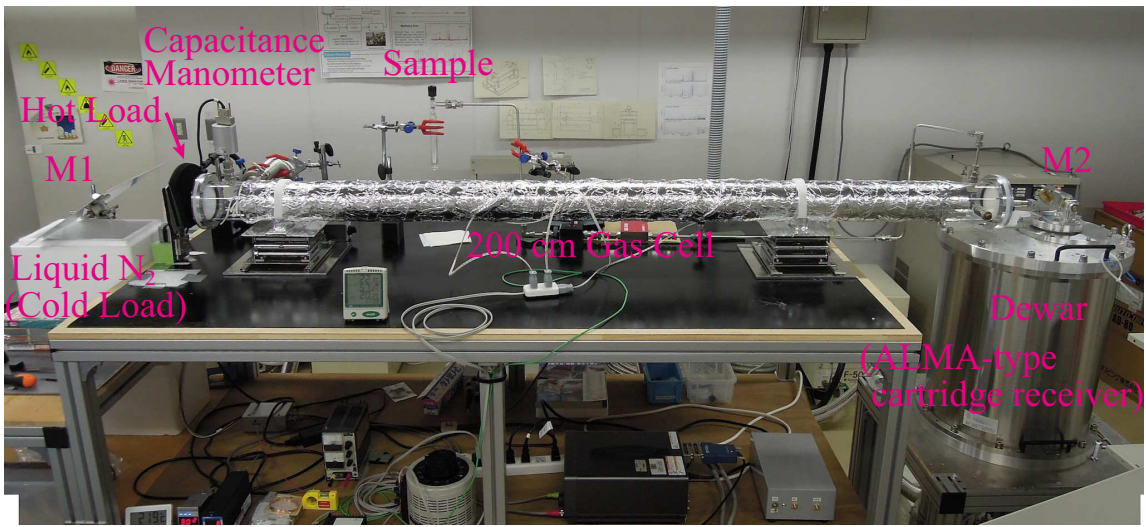}
 \end{center}
 \caption{Side view of the SUMIRE system.}
 \label{fig15}
\end{figure*}

\section{Apparatus}
\subsection{Overview}
A block diagram of an experimental setup and a photograph of the Spectrometer Using superconductor MIxer Receiver (SUMIRE) are shown in Figures~\ref{fig01} and \ref{fig15}, respectively.  An ALMA-type cartridge receiver \citep{Kerr2014} is used to detect emission from molecules in an enclosed glass gas cell against black body radiation at the liquid nitrogen temperature (77~K).  Both sides of the gas cell are sealed by two Teflon plates having a thickness of 10~mm.  In addition, the temperature of the cell can be controlled from a room temperature ($\sim 296$~K) to $\sim 400$~K since the tape heaters are evenly wound around the gas cell.   The emission from molecules in the gas cell is fed into the receiver by a flat mirror (M2).  The emission signal is converted to the intermediate frequency (IF) signal (4-8~GHz) by mixing the first local oscillator signal in the SIS mixers.  The IF signal is further processed to two chunks of the second IF signals 200-2200~MHz by the IF converter units and are introduced to eXtended Fast Fourier Transform Spectrometer (XFFTS) \citep{Klein2006,Klein2012}.  In order to ensure the frequency accuracy, the first and second local oscillators are synchronized with a 10~MHz reference signal generated by a rubidium atomic clock, which gives $2\times 10^{-12}$ or better accuracy over 100 seconds.  Long term variation of the rubidium atomic clock is calibrated by the coordinated universal time provided via GPS.  An internal clock in the XFFTS is also synchronized to the 10~MHz reference signal.  An intensity scale is calibrated to the temperature scale (K) by the chopper wheel method as described in section \ref{sec_measure}.

\subsection{Gas Cell and Vacuum System}
A borosilicate cylindrical gas cell (coefficient of thermal expansion $\alpha = 3.25 \times 10^{-6}$ $^\circ\mathrm{C}^{-1}$) with a length of 200~cm and a diameter of 10~cm is utilized for this experiment.  The cell is evacuated by a turbomolecular vacuum pump (Pferifer HiCube 80 Classic) and residual pressure is $< 2 \times 10^{-4}$~Pa.  The cell internal pressure and the temperature of the gas cell surface are monitored by a capacitance manometer and a thermocouple sensor (K type), respectively.  A canula containing a liquid sample is connected to the gas cell with a metal tube.  When the stopper of the canula is opened, the liquid sample is gasified and flows into the gas cell through the metal tube.  The amount of the gas in the cell is controlled manually with a mass flow controller.  

\subsection{Receiver}
An ALMA-type cartridge receiver is adopted in our system.  The receiver is a 2SB type employing two SIS mixers \citep{Kerr1998}.  By using the two SIS mixers, a RF 90$^{\circ}$ hybrid coupler, and an IF 90$^{\circ}$ hybrid coupler, upper sideband (USB) and lower sideband (LSB) signals are separated in the IF bands and are simultaneously observed.  The frequency range is from 210~GHz to 280~GHz with the IF frequency of 4--8~GHz.  A typical receiver noise temperature is $140\pm 50$~K.  The mixer tips and the mixer blocks designed for the ALMA prototype antenna on the ALMA Test Facility (ATF) site \citep{Asayama2003} are utilized in this experiment.   

\subsection{Backend and Intermediate Frequency Converter}
The digital auto-correlation spectrometer XFFTS \citep{Klein2006,Klein2012} is used as backends.  The XFFTS consists of a 10-bit analog digital converter at a sampling rate of 5~GS/s.  The bandwidth is 2500~MHz with 32768 channels and channel spacing of 76.3~kHz (wide mode).  An equivalent noise bandwidth, which corresponds to the frequency resolution, is 88.5~kHz.  The four XFFTS are used to cover all the IF outputs including both the LSB and USB.  In addition to the 2500~MHz XFFTS, high-frequency-resolution measurements require the XFFTS with the bandwidth of 500 MHz (narrow mode).  The channel number, the channel spacing and the equivalent noise bandwidth are 32768, 15.3~kHz, and 17.7~kHz, respectively.

Since the XFFTS crate has 8 slots each for the wide and narrow mode XFFTSs, the maximum bandwidth of 32~GHz is available in the wide mode with the additional XFFTS boards.  Therefore, when the wider instantaneous-IF-bandwidth (e.g. 4-20 GHz) receiver is available in future, simultaneous measurement of 32~GHz, which covers LSB and USB with the bandwidth of 16~GHz each, is possible. 

\begin{figure}
 \begin{center}
  \includegraphics[width=7cm]{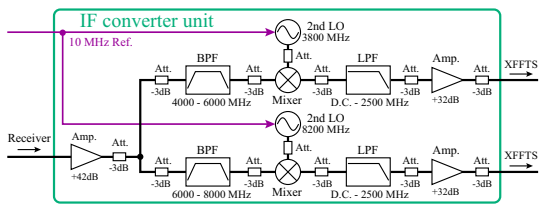}
 \end{center}
 \caption{Block diagram of the IF converter unit.}
 \label{fig02}
\end{figure}

The output IF signal (4-8~GHz) from the receiver is converted to the two chunks of the 200-2200~MHz signal to fit them to the appropriate frequency range (DC-2500~MHz) of the XFFTS by the IF converter unit.  Figure~\ref{fig02} is a block diagram of the IF converter unit.  The IF signal from the receiver is divided into two signals by a power divider.  One of them passes through a bandpass filter of 4000-6000~MHz.  Then, the 4000-6000~MHz signal is converted to be 200-2200~MHz by mixing an LO signal of 3800~MHz.  The other IF signal passes through a bandpass filter of 6000-8000~MHz and is converted to 200-2200~MHz by mixing an LO signal of 8200~MHz.  The 200-2200~MHz signals are introduced into the XFFTS via low-pass filters with a passband of $<2500$~MHz.  When the high-resolution XFFTS is used, the signal from the IF converter unit passes through a low-pass filter with pass band of $<400$~MHz before going into the 500~MHz XFFTS to prevent aliasing of the signal. 

\subsection{Intensity Calibration \label{sec_measure}}
An emission spectrum of the gas sample is measured against the cold load cooled down to 77~K by liquid nitrogen.  In addition, blank spectra against the cold load and the hot load at the room temperature are measured for intensity calibration.  Hence, a single spectroscopic measurement consists of these three measurements.  Identical integration time is applied to each measurement, which is set to be 5~minutes (referred hereafter as the integration time) unless otherwise stated.  In this case, it takes 15~minutes plus overhead time to obtain a single spectrum. 

By assuming the Rayleigh-Jeans law, the measured power at the backend can be represented on a temperature scale, as employed in radio-astronomical observations.  When the molecular emission is measured against the cold load, the output power is represented as:
\begin{equation}
S_{\rm mol} = k (T_{\rm mol} + T_{\rm cold} + T_{\rm sys}),  \\
\end{equation}
where $T_{\rm mol}$ is the intensity of the molecular emission on a temperature scale, $T_{\rm cold}$ the cold load temperature, and $T_{\rm sys}$ the system noise temperature. The coefficient $k$ is a proportionality coefficient specific to the instrument. Similarly, the output power observing the hot and cold loads with a blank cell is given as:
\begin{equation}
S_{\rm hot} = k (T_{\rm hot} + T_{\rm sys}),  \\
\end{equation}
and
\begin{equation}
S_{\rm cold} = k (T_{\rm cold} + T_{\rm sys}),  \\
\end{equation}
where $T_{\rm hot}$ is the hot load temperature (i.e., the room temperature).  Then, the intensity of the molecular emission on a temperature scale is calculated from $S_{\rm mol}$, $S_{\rm hot}$, and $S_{\rm cold}$ as:  
\begin{equation}
\label{eq4}
T_{\rm mol} = \frac{S_{\rm mol} - S_{\rm cold}}{S_{\rm hot} - S_{\rm cold}} \times (T_{\rm hot} - T_{\rm cold}).
\end{equation}
In this way, the molecular emission spectrum on an absolute intensity scale can be obtained.  All these measurements are controlled by using PC, while the handling of the gas sample is manually performed.  In general, repeating the spectroscopic measurements lead to achieving an appropriate signal-to-noise (S/N) ratio.  

The calibrated spectrum is further reduced by using the software package \textit{CLASS}\footnote{http://www.iram.fr/IRAMFR/GILDAS} developed by institut de radioastronomie millim\'{e}trique (IRAM).  The spectrum obtained by using equation (\ref{eq4}) suffers from baseline ripples caused by standing waves in the optical path.  Hence, the baseline is subtracted by using the 15th-20th order polynomial function to the line free part of about 200~MHz range.  Since the typical linewidth of the spectrum is as narrow as 300~kHz in the 200~GHz band at the room temperature and the sample pressure of 0.5~Pa, the baseline subtraction by the 20th order polynomial function does not affect spectral line shapes. 

\section{Results}
Test measurements of SUMIRE were carried out by making use of CH$_3$CN, HDO, and CH$_3$OH. Results for each species are described below. 

\begin{figure*}[t]
 \begin{center}
  \includegraphics[width=15cm]{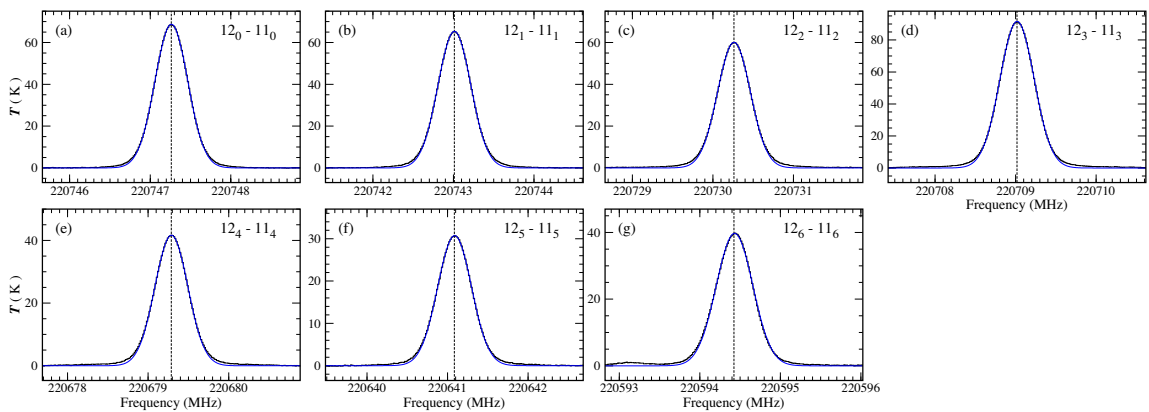}
 \end{center}
 \caption{Spectra of (a) CH$_3$CN ($12_0-11_0$), (b) CH$_3$CN ($12_1-11_1$), (c) CH$_3$CN ($12_2-11_2$), (d) CH$_3$CN ($12_3-11_3$), (e) CH$_3$CN ($12_4-11_4$), (f) CH$_3$CN ($12_5-11_5$), and (g) CH$_3$CN ($12_6-11_0$) measured with the SUMIRE.  The vertical dashed lines are the frequencies cataloged in CDMS (also see Table~\ref{tab02}).  The blue lines are the results of Gaussian fitting, where the central part of the spectrum with the intensity higher than $1/e$ of the peak intensity is used for the fitting to avoid a remaining contribution of the Lorenzian skirt.  The spectra are measured at the gas pressure of 0.05-0.09~Pa with the total integration time of 40~min.  The frequency resolution is 17.7~kHz.}
 \label{fig03}
\end{figure*}

\begin{table*}[t]
\tbl{Measured frequencies of CH$_3$CN. }{
\small
\begin{tabular}{llcll}
\hline \hline
 Transition & Freq. (CDMS) & XFFTS$^{\rm a}$ & Freq. (SUMIRE) $^{\rm b,c}$ & SUMIRE - CDMS$^{\rm d}$ \\
            & (MHz)        &                 & (MHz)              & (MHz)                   \\
\hline \hline
 $12_0 - 11_0,\, v=0$ & 220747.2617 (2) & n & 220747.2628 (1)  & 0.0011  \\
                      &                 & w & 220747.2631 (2)  & 0.0014  \\ 
 $12_1 - 11_1,\, v=0$ & 220743.0111 (2) & n & 220743.0121 (1)  & 0.0010  \\
                      &                 & w & 220743.0124 (2)  & 0.0013  \\ 
 $12_2 - 11_2,\, v=0$ & 220730.2611 (2) & n & 220730.2619 (1)  & 0.0008  \\
                      &                 & w & 220730.2627 (2)  & 0.0016  \\ 
 $12_3 - 11_3,\, v=0$ & 220709.0170 (2) & n & 220709.0181 (2)  & 0.0011  \\
                      &                 & w & 220709.0177 (4)  & 0.0007  \\ 
 $12_4 - 11_4,\, v=0$ & 220679.2874 (2) & n & 220679.2889 (1)  & 0.0015  \\
                      &                 & w & 220679.2894 (1)  & 0.0020  \\ 
 $12_5 - 11_5,\, v=0$ & 220641.0844 (2) & n & 220641.0878 (2)  & 0.0034  \\
                      &                 & w & 220641.0869 (4)  & 0.0025  \\ 
 $12_6 - 11_6,\, v=0$ & 220594.4237 (2) & n & 220594.4314 (5)  & 0.0077  \\
                      &                 & w & 220594.4305 (13) & 0.0068  \\
 $13_0 - 12_0,\, v=0$ & 239137.9168 (2) & n & 239137.9176 (1)  & 0.0008  \\
                      &                 & w & 239137.9176 (2)  & 0.0008  \\
 $13_1 - 12_1,\, v=0$ & 239133.3134 (2) & n & 239133.3141 (1)  & 0.0007  \\
                      &                 & w & 239133.3141 (2)  & 0.0007  \\
 $13_2 - 12_2,\, v=0$ & 239119.5049 (2) & n & 239119.5058 (1)  & 0.0009  \\
                      &                 & w & 239119.5059 (1)  & 0.0010  \\
 $13_3 - 12_3,\, v=0$ & 239096.4971 (2) & n & 239096.4981 (2)  & 0.0010  \\
                      &                 & w & 239096.4982 (5)  & 0.0011  \\
 $13_4 - 12_4,\, v=0$ & 239064.2993 (2) & n & 239064.3008 (1)  & 0.0015  \\
                      &                 & w & 239064.3009 (2)  & 0.0016  \\
 $13_5 - 12_5,\, v=0$ & 239022.9246 (2) & n & 239022.9267 (2)  & 0.0021  \\
                      &                 & w & 239022.9267 (2)  & 0.0021  \\
 $13_6 - 12_6,\, v=0$ & 238972.3900 (2) & n & 238972.3939 (2)  & 0.0039  \\
                      &                 & w & 238972.3933 (3)  & 0.0033  \\
\hline
\end{tabular}}\label{tab02}
\begin{tabnote}
\footnotemark[a] n and w indicate 500~MHz XFFTS and 2500~MHz XFFTS, respectively. \\
\footnotemark[b] The numbers in parentheses represent $1\sigma$ error. \\
\footnotemark[c] The frequencies derived by the Gaussian fitting to the averaged spectrum. \\
\footnotemark[d] The difference between the frequency of SUMIRE and that of CDMS. \\
\end{tabnote}
\end{table*}

\subsection{CH$_3$CN \label{sec_ch3cn}}
For CH$_3$CN, 7 $K$-structure lines ($K=0-6$) each for $J=12-11$ and $13-12$ transitions in the ground vibrational state were measured in the high frequency-resolution mode by using the 500~MHz XFFTS.  The measurements were performed at the room temperature and the sample pressure of 0.05-0.09~Pa.  The every set of measurements (consisting of the molecular emission, cold load, and hot load measurements) is repeated 20 times with the integration time of 2 minutes. 

The averaged CH$_3$CN spectra of $J=12-11$ transitions derived from the repeated measurements are shown in Figure~\ref{fig03}.  The frequencies are determined by fitting the Gaussian function to the averaged spectra.  Here, the Gaussian function was employed for the spectral line shape function because the Doppler effect due to the Maxwell-Boltzmann distribution makes a dominant contribution to the line shape in the measured frequency range.  Although a Lorentzian component seems to remain in the skirt of the spectral lines, as shown in Figure~\ref{fig03}, it does not affect the frequency measurement.  In fact, the frequencies derived with the Gaussian function are confirmed to be consistent with those determined with the pseudo-Voigt function within $1 \sigma$ error.  The results are summarized in Table~\ref{tab02}.  

The frequencies determined with SUMIRE are compared with those listed in CDMS, as shown in Table~\ref{tab02}. The frequency with SUMIRE are systematically higher by 0.7-7.7~kHz than those in CDMS. This is partly due to the hyperfine splitting by the nuclear quadrupole interaction of the $^{14}$N nucleus.  The hyperfine structure is too small to be resolved (typically from 15~kHz to 200~kHz for $12_0-11_0$ and $12_6-11_6$, respectively), but it gives slight asymmetry in the unresolved spectral line profile, which makes a slight shift of the line centroid from the hypothetical center frequency without the hyperfine interaction listed in CDMS.  Since the hyperfine splitting is larger for higher $K$ lines, the frequency shift is expected to be larger for them.  We simulate an unresolved spectral profile considering the hyperfine components each of which has a Gaussian profile and fit it with a single Gaussian function.  Then, we confirm that the line centroid obtained by the single Gaussian fitting tends to be shifted to higher frequency for higher $K$ lines.  Indeed, the observed $K=5$ and $K=6$ lines show larger deviations than those of $K=0$, 1, and 2 lines by a factor of 5-7.  On the other hand, the effect of the hyperfine splitting is almost negligible for the lower $K$ lines (i.e., $K=0$, 1, and 2).  For example, the centroid of $12_0-11_0$ line is evaluated to be 220747.269~MHz by the Gaussian fitting considering the hyperfine components and is almost the same as 220747.2628~MHz obtained by the single Gaussian fitting.  Thus, the frequency shifts observed for these lines cannot be explained by this mechanism.  This fact means that a small systematic difference ($\sim 8$~kHz at most) remains between the frequencies with SUMIRE and those in CDMS.  Nevertheless, the difference for the lower $K$ lines ($\sim 1$~kHz) is much smaller than the frequency resolution of the digital spectrometer, and the absolute frequency accuracy of our system is guaranteed by the rubidium clock calibrated by the GPS system.  Hence, we conclude that SUMIRE can provide the frequency as accurate as $\sim 1$~kHz for lines measured with a high S/N ratio.  It is also revealed that SUMIRE is sensitive to small frequency shifts caused by unresolved hyperfine components. 

In addition to the measurements with the 500~MHz XFFTS, the measurements of CH$_3$CN were also carried out with the 2500~MHz XFFTS.  As with the 500~MHz XFFTS measurements, the frequencies of the CH$_3$CN lines are determined with the single Gaussian function fitting and are summarized in Table~\ref{tab02}.  The frequencies measured with the 2500~MHz XFFTS are confirmed to be consistent with those measured with the 500~MHz XFFTS within 1~kHz which is smaller than the frequency resolution of 2500~MHz XFFTS (88.5~kHz) by a factor of 90.  This result assures that the 2500~MHz XFFTS can determine the frequency with precision of $\sim 1$~kHz for strong emission lines.

\begin{figure*}[t]
 \begin{center}
  \includegraphics[width=15cm]{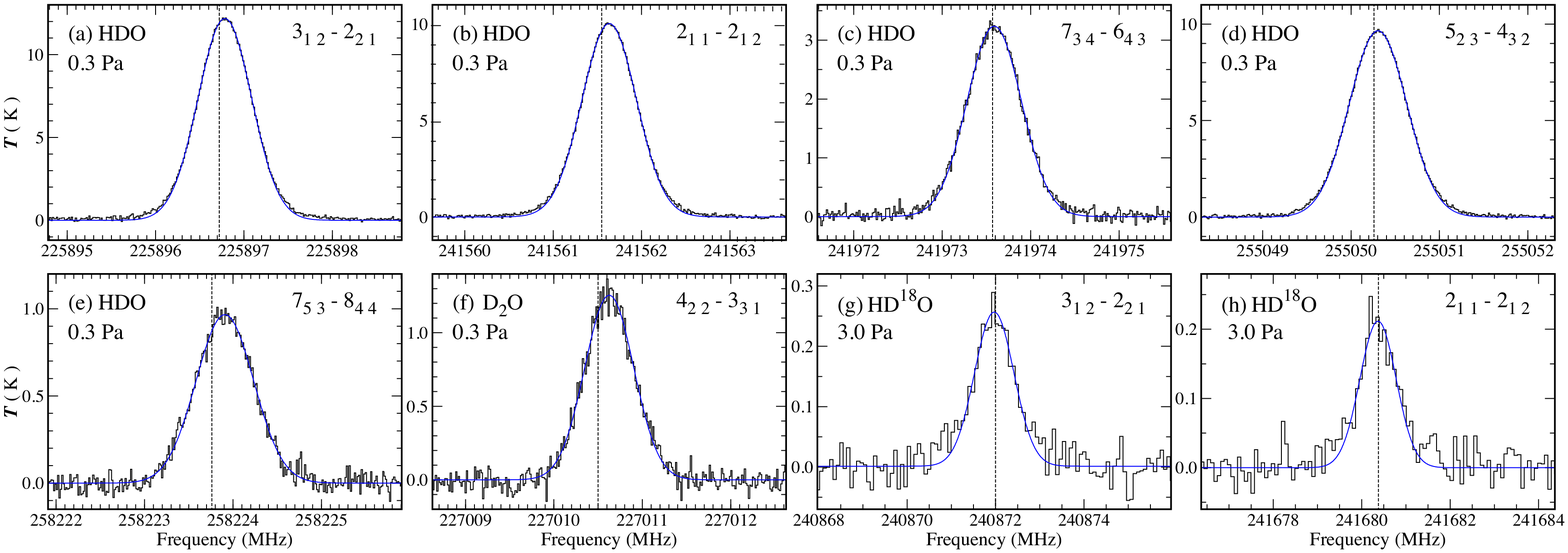}
 \end{center}
 \caption{Spectra of (a) HDO ($3_{1\,2} - 2_{2\,1}$), (b) HDO ($2_{1\,1} - 2_{1\,2}$), (c) HDO ($7_{3\,4} - 6_{4\,3}$), (d) HDO ($5_{2\,3} - 4_{3\,2}$), (e) HDO ($7_{5\,3} - 8_{4\,4}$), (f) D$_2$O ($4_{2\,2} - 3_{3\,1}$), (g) HD$^{18}$O ($3_{1\,2} - 2_{2\,1}$), and (h) HD$^{18}$O ($2_{1\,1} - 2_{1\,2}$) measured with the SUMIRE.  The vertical dashed lines are the frequencies cataloged in JPL (also see Table~\ref{tab04}).  The blue lines are the results of Gaussian fitting, where the central part of the spectrum with the intensity higher than $1/e$ of the peak intensity is used for the fitting to avoid a remaining contribution of the Lorenzian skirt.  The spectra are measured at the gas pressure of 0.3~Pa with the total integration time of 50~min at the frequency resolution of 17.7~kHz for HDO and D$_2$O, while the spectra are measured at the gas pressure of 3~Pa with the total integration time of 110~min at the frequency resolution of 88.5~kHz for HD$^{18}$O. }
 \label{fig04}
\end{figure*}

\begin{table*}[t]
\tbl{Measured frequencies of HDO, D$_2$O, and HD$^{18}$O. }{
\small
\begin{tabular}{lllll}
\hline \hline
Mol. & Transition & Freq. (JPL) & Freq. (SUMIRE)$^{\rm a,b}$  & SUMIRE - JPL$^{\rm f}$ \\
     &            & (MHz)       & (MHz)                       & (MHz)                  \\
\hline \hline
HDO        & $3_{1\,2} - 2_{2\,1}$ & 225896.720 (38)   & 225896.7840 (5)  & 0.064      \\
HDO        & $2_{1\,1} - 2_{1\,2}$ & 241561.550 (37)   & 241561.6341 (5)  & 0.084      \\
HDO        & $7_{3\,4} - 6_{4\,3}$ & 241973.570 (39)   & 241973.5883 (19) & 0.018      \\
HDO        & $5_{2\,3} - 4_{3\,2}$ & 255050.260 (59)   & 255050.3051 (5)  & 0.045      \\
HDO        & $7_{5\,3} - 8_{4\,4}$ & 258223.760 (104)  & 258223.9080 (29) & 0.148      \\
D$_2$O     & $4_{2\,2} - 3_{3\,1}$ & 227010.500 (90)   & 227010.6218 (41) & 0.122      \\
HD$^{18}$O & $3_{1\,2} - 2_{2\,1}$ & 240872.00 (20)    & 240871.970  (12) & -0.03      \\
HD$^{18}$O & $2_{1\,1} - 2_{1\,2}$ & 241680.38 (20)    & 241680.370  (21) & -0.01      \\
\hline
\end{tabular}}\label{tab04}
\begin{tabnote}
\footnotemark[a] The numbers in parentheses represent $1\sigma$ error. \\
\footnotemark[b] The frequencies derived by the Gaussian fitting to the averaged spectrum. \\
\footnotemark[c] The difference between the frequency of SUMIRE and that of JPL. \\
\end{tabnote}
\end{table*}
\subsection{HDO and HD$^{18}$O}
Five rotational transitions of HDO were measured at the room temperature by using the 500~MHz XFFTS.  The spectroscopic measurements were repeated 10 times with the integration time of 5 minutes.  A mixture of the equal amount of liquid H$_2$O and liquid D$_2$O is prepared in a sample cylinder, and its vapor is introduced into the gas cell of SUMIRE.  The total pressure was $\sim0.3$~Pa, which means the partial pressure of HDO of $\sim 0.15$~Pa. 

Figure~\ref{fig04} shows the spectral lines of the five HDO and one D$_2$O transitions averaged over all the measurements.  The transition frequency was obtained by fitting the Gaussian function to the averaged spectrum for each transition.  The results are listed in Table~\ref{tab04}.  As mentioned in Section \ref{sec_ch3cn}, the Doppler width is dominant, and hence, the use of the Gaussian function is justified.  Indeed, the full width at half maximum (FWHM) of the linewidth is derived to be $0.85-0.97$~km~s$^{-1}$ for HDO and 0.87~km~s$^{-1}$ for D$_2$O which are almost comparable to the Doppler widths of 0.85~km~s$^{-1}$ and 0.83~km~s$^{-1}$ at 300~K, respectively.  

Furthermore, two transitions of HD$^{18}$O were measured by using the 2500~MHz XFFTS in this study.  In this case, the sample pressure was increased up to 3~Pa, and the measurement was repeated 22 times to increase the S/N ratio.  Figure~\ref{fig04} represents the observed spectra of HD$^{18}$O.  The transition frequencies were determined by fitting the Gaussian function to the spectra averaged over the 22 measurements, and are listed in Table~\ref{tab04}.  The FWHM of the HD$^{18}$O linewidth was evaluated to be $1.3-1.5$~km~s$^{-1}$, which is significantly broader than the Doppler width (0.81~km~s$^{-1}$).  The sample pressure is 10 times higher than that in the HDO and D$_2$O measurements, and hence, the pressure broadening is dominant.  Even in this case, the Gaussian fit to the central part of the line still works for evaluating the center frequency (See caption of Figure~\ref{fig04}).

\begin{figure*}[t]
 \begin{center}
  \includegraphics[width=15cm]{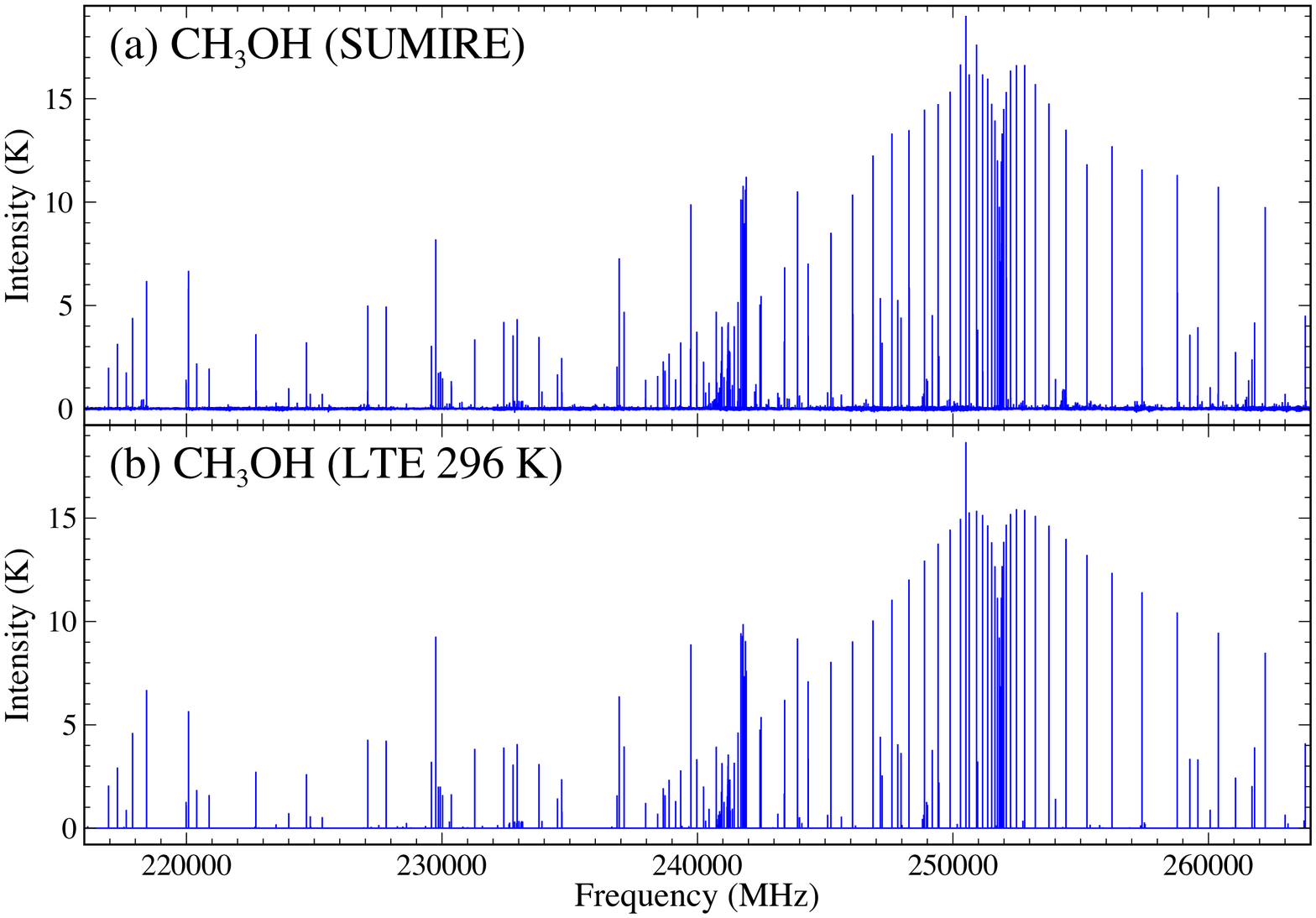}
 \end{center}
 \caption{(a) CH$_3$OH spectrum obtained by the SUMIRE and (b) a spectrum calculated under the assumption of the LTE condition at 296~K.} 
 \label{fig05}
\end{figure*}

\begin{figure*}[t]
 \begin{center}
  \includegraphics[width=15cm]{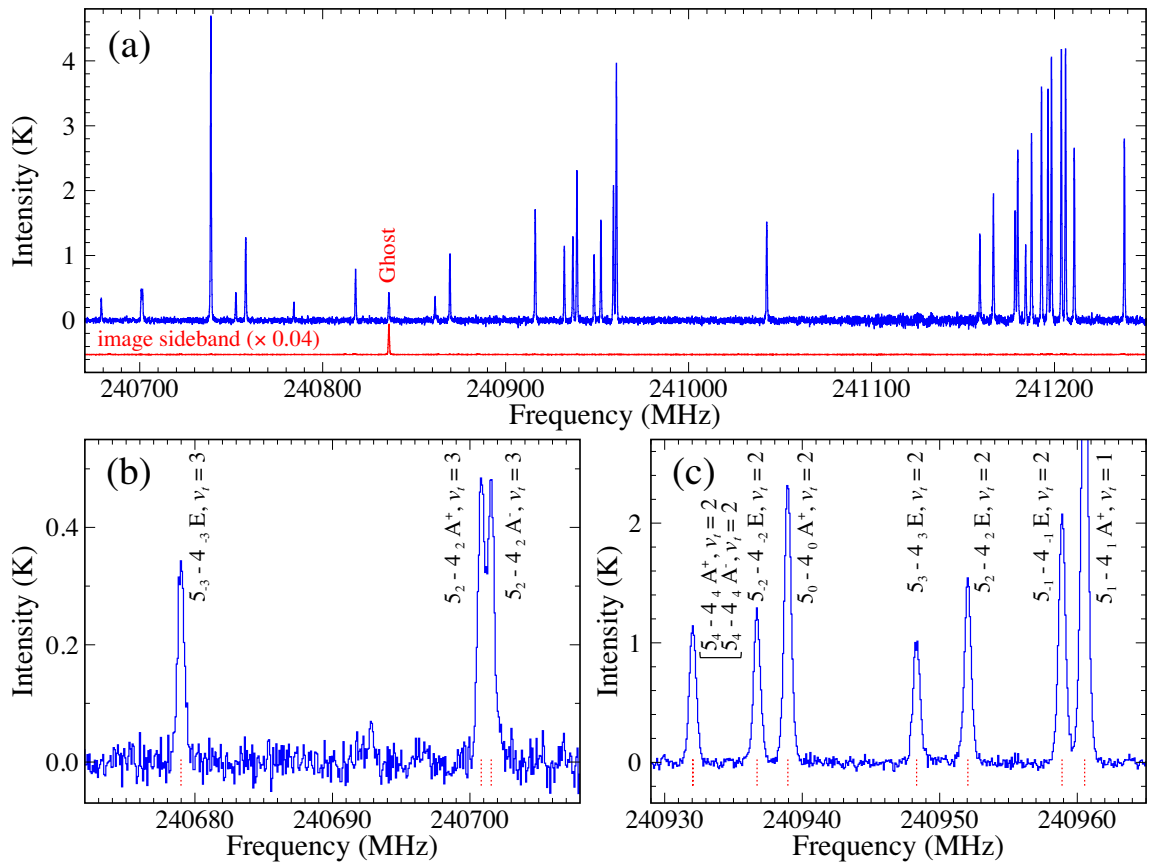}
 \end{center}
 \caption{(a) An expanded spectrum of CH$_3$OH from 240670.0~MHz to 241250.0~MHz, (b) an expanded spectrum of CH$_3$OH from 240672.0~MHz to 240708.0~MHz, and (c) an expanded spectrum of CH$_3$OH from 240926.0~MHz to 240965.0~MHz.  A red spectrum of (a) is an image sideband spectrum of the frequency range.  Ghost indicates an emission line contaminated from the image sideband.  Red dashed vertical lines in (b) and (c) indicate positions of frequency determined by the Gaussian fittings.} 
 \label{fig06}
\end{figure*}

\begin{table*}
\tbl{Measurement parameters for CH$_3$OH. }{
\small
\begin{tabular}{lllllllll}
\hline \hline
ID & 1st LO $^{\rm a}$ & SB$^{\rm b}$ & 2nd LO  & Freq. res.$^{\rm c}$ & Frequency range$^{\rm d}$ & Pres.$^{\rm e}$ & Int.$^{\rm f}$ & No.$^{\rm g}$ \\
   & (MHz)             &              & (MHz)   & (kHz)              & (MHz)                     & (Pa)            & (min)          &          \\
\hline \hline
1  & 224000.000 & L & 8200.000  & 88.5          & 216000.000 - 218000.000 & 0.46-0.50    & 5         & 20 (10)  \\
   & ...        & L & 3800.000  & ...           & 218000.000 - 220000.000 & ...          & ...       & ...      \\
   & ...        & U & 3800.000  & ...           & 228000.000 - 230000.000 & ...          & ...       & ...      \\
   & ...        & U & 8200.000  & ...           & 230000.000 - 232000.000 & ...          & ...       & ...      \\
2  & 228000.000 & L & 8200.000  & 88.5          & 220000.000 - 222000.000 & 0.45-0.51    & 5         & 20 (10)  \\
   & ...        & L & 3800.000  & ...           & 222000.000 - 224000.000 & ...          & ...       & ...      \\
   & ...        & U & 3800.000  & ...           & 232000.000 - 234000.000 & ...          & ...       & ...      \\
   & ...        & U & 8200.000  & ...           & 234000.000 - 236000.000 & ...          & ...       & ...      \\
3  & 232000.000 & L & 8200.000  & 88.5          & 224000.000 - 226000.000 & 0.46-0.50    & 5         & 20 (10)  \\
   & ...        & L & 3800.000  & ...           & 226000.000 - 228000.000 & ...          & ...       & ...      \\
   & ...        & U & 3800.000  & ...           & 236000.000 - 238000.000 & ...          & ...       & ...      \\
   & ...        & U & 8200.000  & ...           & 238000.000 - 240000.000 & ...          & ...       & ...      \\
4  & 248000.000 & L & 8200.000  & 88.5          & 240000.000 - 242000.000 & 0.46-0.51    & 5         & 20 (10)  \\
   & ...        & L & 3800.000  & ...           & 242000.000 - 244000.000 & ...          & ...       & ...      \\
   & ...        & U & 3800.000  & ...           & 252000.000 - 254000.000 & ...          & ...       & ...      \\
   & ...        & U & 8200.000  & ...           & 254000.000 - 256000.000 & ...          & ...       & ...      \\
5  & 252000.000 & L & 8200.000  & 88.5          & 244000.000 - 246000.000 & 0.46-0.51    & 5         & 20 (10)  \\
   & ...        & L & 3800.000  & ...           & 246000.000 - 248000.000 & ...          & ...       & ...      \\
   & ...        & U & 3800.000  & ...           & 256000.000 - 258000.000 & ...          & ...       & ...      \\
   & ...        & U & 8200.000  & ...           & 258000.000 - 260000.000 & ...          & ...       & ...      \\
6  & 256000.000 & L & 8200.000  & 88.5          & 248000.000 - 250000.000 & 0.45-0.50    & 5         & 20 (10)  \\
   & ...        & L & 3800.000  & ...           & 250000.000 - 252000.000 & ...          & ...       & ...      \\
   & ...        & U & 3800.000  & ...           & 260000.000 - 262000.000 & ...          & ...       & ...      \\
   & ...        & U & 8200.000  & ...           & 262000.000 - 264000.000 & ...          & ...       & ...      \\
\hline
\end{tabular}}\label{tab05}
\begin{tabnote}
\footnotemark[a] Measurements with the LO frequency shifted by $+39.0625$~MHz are also conducted to identify spurious lines such as contamination from the image sideband and harmonic mixing in the IF converter unit.\\
\footnotemark[b] The sideband of receiver.  L and U indicate LSB and USB, respectively. \\
\footnotemark[c] The frequency resolution of the spectrometer XFFTS.  The channel spacing is 88.5~kHz. \\
\footnotemark[d] The frequency range covered by the spectrometer.  \\
\footnotemark[e] The pressure of the gas cell.  \\
\footnotemark[f] The integration time of each scan in minutes.  \\
\footnotemark[g] The total number of measurements.  The number in the parentheses is the number of measurements with the 1st LO frequency shifted by $+39.0625$~MHz. \\
\end{tabnote}
\end{table*}

\subsection{CH$_3$OH}
We measured the CH$_3$OH transitions in a wide frequency range from 216~GHz to 264~GHz.  Six frequency settings shown in Table~\ref{tab06} were applied to cover this range.  Each frequency setting covers the frequency range of 8~GHz by using the four 2500~MHz XFFTSs.  The measurements were conducted at the room temperature and the gas sample pressure of $\sim0.5$~Pa with the total integration time of 50~minutes (10 times the 5 min integration).  In addition, the same measurements were carried out with the first LO frequency shifted by $+39.0625$~MHz.  By comparing the two spectra measured with the slightly different first LO frequencies, we examined ghost emission lines caused by contaminations from the image sideband due to insufficient sideband rejection of the 2SB receiver and by a harmonic mixing in the frequency conversion in the IF unit.  After confirmation of ghost emission lines, the two spectra were averaged to obtain the final spectrum.  An overall view of the spectrum is shown in Figure~\ref{fig05}.  

Figure~\ref{fig06} shows an example of the expanded spectra of CH$_3$OH.  The line shapes are well resolved with the frequency resolution of 88.5~kHz.  In total, 295 emission lines are detected with the S/N ratio higher than $>3$, excluding the ghost lines.  From the 274 detected emission lines, we identified 242 transitions of CH$_3$OH, including 16 transitions of the third torsionally excited state of CH$_3$OH ($v_t=3$), and 51 transitions of $^{13}$CH$_3$OH with aid of the data from the CDMS \citep{muller01,cdms2005} and \citet{Pearson2009}.  Here, 19 emission lines out of the 274 identified emission lines consist of two transitions of which frequencies are too close to separate by the Gaussian fitting.  After the spectral assignments, 21 unidentified emission lines remain.  The line parameters obtained by the Gaussian fittings are summarized in Appendix.  The FWHM of CH$_3$OH linewidths are evaluated to be 0.5-1.2~km~s$^{-1}$ by the Gaussian fitting.  These values tend to be slightly larger than the Doppler width of 0.66~km~s$^{-1}$ at 300~K.  This result indicates that the pressure broadening is thought to contribute to the linewidth of CH$_3$OH.

A histogram of the number of CH$_3$OH lines as a function of frequency difference between our measurements and the data from the CDMS is shown in Figure~\ref{fig14}.  55 out of 242 emission lines are utilized in this histogram and satisfy two criteria: (1) the uncertainty of frequency given by the CDMS data is less than 10~kHz, and (2) that of this measurement is less than 5~kHz.  About 50~\% of frequencies derived from the measurements reasonably correspond with those in the CDMS data within $\pm 10$~kHz.  On the other hand, the rest 50~\% of the frequencies are found to deviate by 10~kHz or larger.  The peak position of the histogram seems to be situated between 0~kHz and $-20$~kHz.

\begin{figure}
 \begin{center}
  \includegraphics[width=6.5cm]{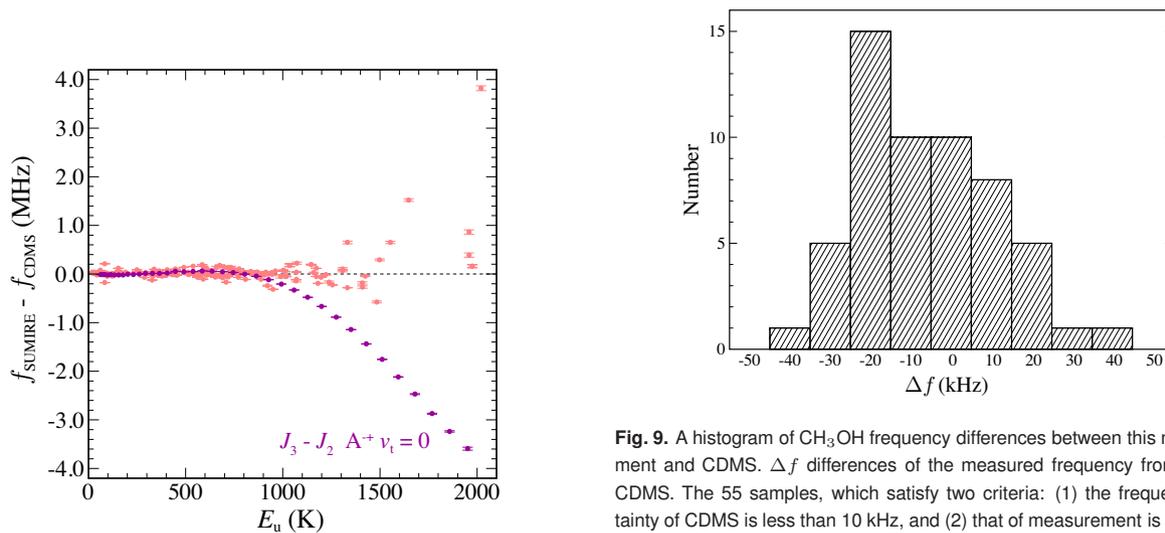}
 \end{center}
 \caption{Differences of CH$_3$OH frequencies between our measurement ($f_{\rm SUMIRE}$) and CDMS ($f_{\rm CDMS}$) as a function of the upper state energy.  Purple plots indicate a series of the Q-branch ($J_{3}-J_{2}$ A$^{-+}\,v_{\rm t} = 0$), where $J$ is the rotational quantum number. } 
 \label{fig07}
\end{figure}

\begin{figure}[t]
 \begin{center}
  \includegraphics[width=6.5cm]{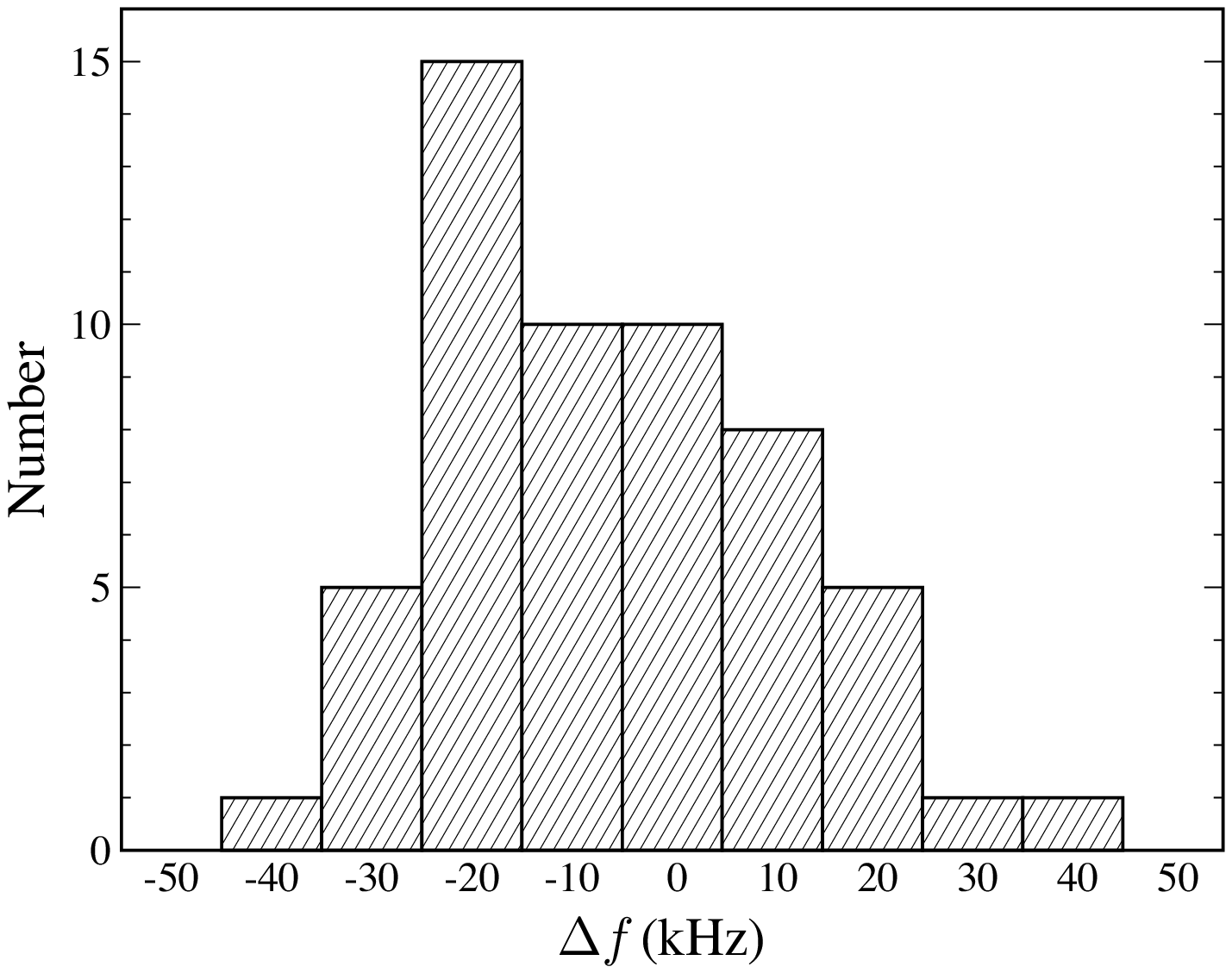}
 \end{center}
 \caption{A histogram of CH$_3$OH frequency differences between this measurement and CDMS.  $\Delta f$ differences of the measured frequency from that of CDMS.  The 55 samples, which satisfy two criteria: (1) the frequency certainty of CDMS is less than 10~kHz, and (2) that of measurement is less than 5~kHz, are used for the histogram.  } 
 \label{fig14}
\end{figure}

Figure~\ref{fig07} shows CH$_3$OH frequency differences between the measurement and the data from the CDMS as a function of the upper state energy $E_{\rm u}$.  Here, all the transition frequencies are used in the plot, except for the blended spectral lines and the lines in the $v_t=3$ state.  For the lines with $E_{\rm u}<1000$~K, most of frequencies accord with each other within $\sim 100$~kHz, while significant differences are found for the lines with $E_{\rm u}>1000$~K.  For example, the Q-branch series lines with $J_3-J_2 \, {\rm A}^{-+}$, where $J$ is the total rotational quantum number, show a systematic deviation as increasing $E_u$.  CH$_3$OH is a floppy molecule with an internal rotation, and the vibration-rotation interaction for this molecule is much more complicated than that for `rigid' molecules.  For this reason, sophisticated Hamiltonian with many interaction parameters is employed for the analysis of the CH$_3$OH spectrum (e.g., \cite{Xu2008}).  However, spectral lines cannot fully be reproduced even with the best effort. Since the frequencies listed in the CDMS are based on the analysis of low excitation lines, ignorance of the higher-order Hamiltonian terms not included in the analysis could cause the above systematic deviation. 

\section{Evaluations of System}
\subsection{Sensitivity}
\begin{figure}
 \begin{center}
  \includegraphics[width=6.5cm]{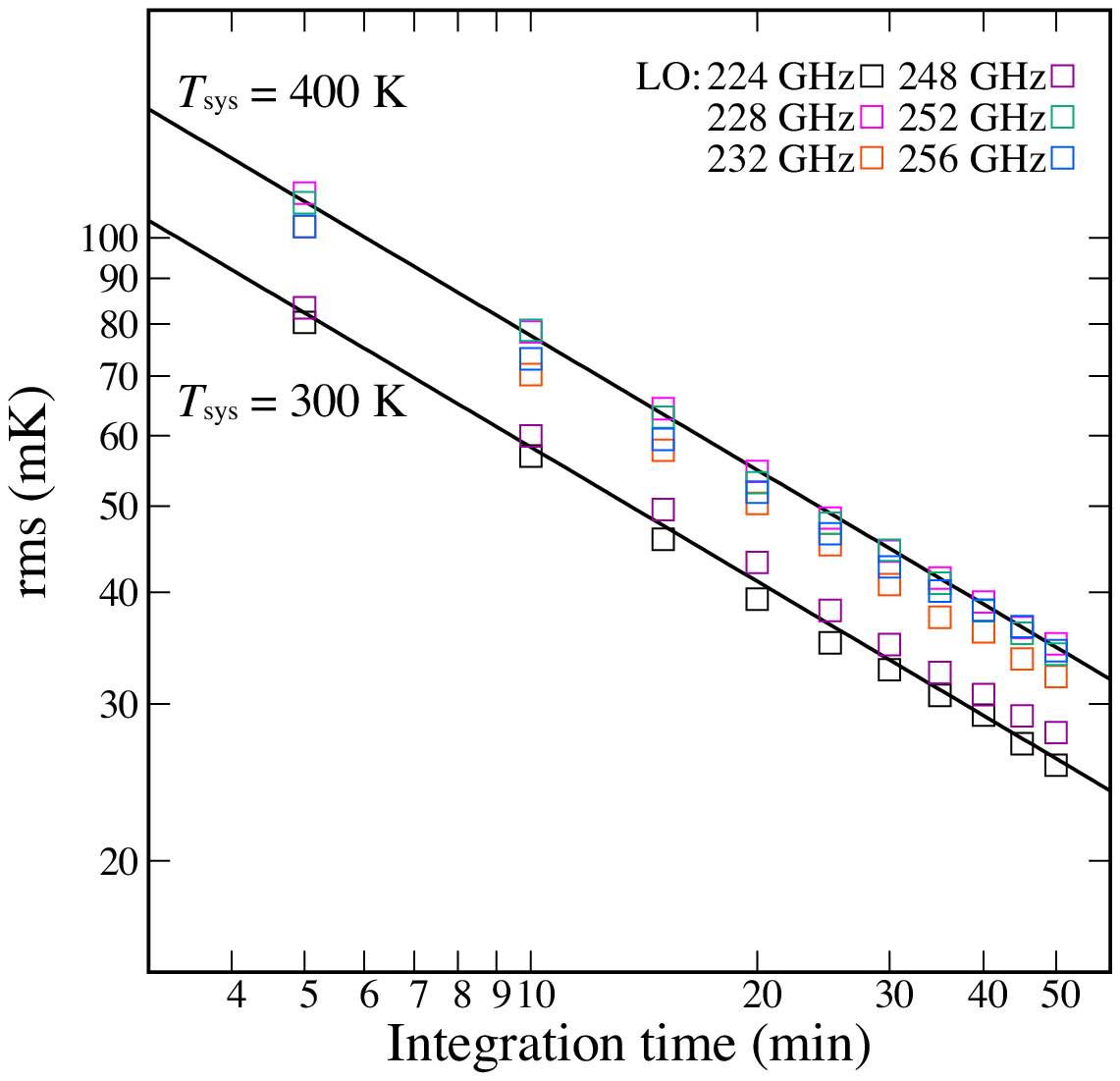}
 \end{center}
 \caption{The rms noise temperatures of CH$_3$OH measurements in the USB as a function of the integration time.  Black lines indicate the ideal rms noise temperature with the system noise temperature of 300~K and 400~K by using equation (\ref{eq7}).  LO indicates the 1st LO frequency of each frequency settings.}
 \label{fig08}
\end{figure}

The rms noise of the measurement in the temperature unit is represented as:
\begin{equation}
\label{eq7}
T_{\rm rms} = \frac{\sqrt{2} T_{\rm sys}}{\sqrt{B \tau}},
\end{equation}
where $T_{\rm sys}$ is the system noise temperature, $B$ is the bandwidth of a frequency channel, and $\tau$ is the integration time.  In order to confirm that the rms noise decreases with the integration time as expected in equation (\ref{eq7}), we evaluate the rms noise of the CH$_3$OH measurements as a function of the integration time ($\tau$).  Figure~\ref{fig08} shows the rms noise measured at different integration time, where the rms noise is evaluated by using a line-free part of the spectrum in the USB for the six frequency settings listed in Table~\ref{tab05}.  The two lines indicate the ideal rms noise calculated by equation (\ref{eq7}) for the system noise temperature of 300~K and 400~K, and a bandwidth of 88.5~kHz, as an eye guide.  These system noise temperatures are typical ones during the CH$_3$OH measurement runs.  Figure~\ref{fig08} indicates that the evaluated rms noise temperature indeed decreases as $\tau^{-\frac{1}{2}}$ up to 50~min. 

With the current system, the system noise temperature (300--400~K) is higher than the receiver noise temperature ($140 \pm 50$~K).  The difference mainly originates from reflection on the Teflon plates and the spillover of the receiver beam outside the cell.  Hence, we can improve the system temperature by replacing the Teflon plates by appropriately designed lenses with antireflection grooves, which is now in progress. With this improvement, we expect to achieve the system noise temperature of SUMIRE of about 100~K: namely, the rms noise after the 5 minute integration will be reduced from 80-100~mK to about 30~mK.  Furthermore, a simultaneous observation of two orthogonal polarizations with expansion of the XFFTSs can contribute to further improvement by a factor of 1.4. 

\subsection{Absolute Intensity}

\begin{figure}[t]
 \begin{center}
  \includegraphics[width=6.5cm]{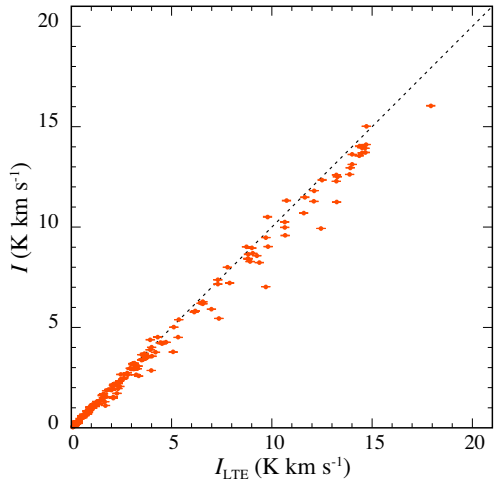}
 \end{center}
 \caption{Measured integrated intensities of CH$_3$OH ($I$) as a function of the integrated intensities estimated by using the LTE approximation ($I_{\rm LTE}$).  A dashed line indicates a ratio of 1. }
 \label{fig09}
\end{figure}

\begin{figure}
 \begin{center}
  \includegraphics[width=6.5cm]{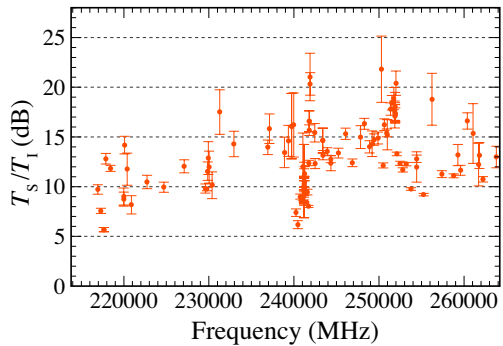}
 \end{center}
 \caption{The $T_{\rm S}/T_{\rm I}$ ratio as a function of the measurement frequency. }
 \label{fig10}
\end{figure}

\begin{figure}[t]
 \begin{center}
  \includegraphics[width=6.5cm]{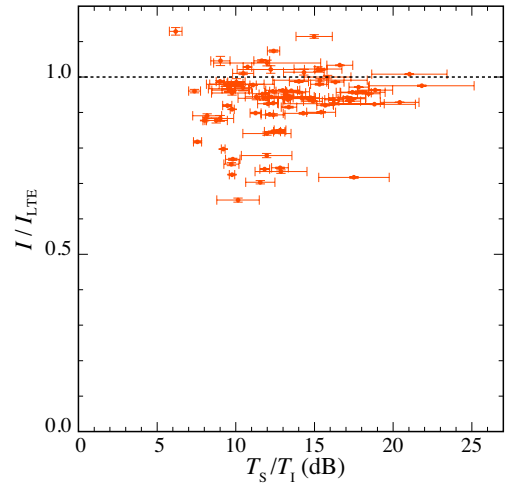}
 \end{center}
 \caption{Ratios of the measure integrated intensity of CH$_3$OH and those evaluated by the LTE approximation ($I/I_{\rm LTE}$) as a function of the $T_{\rm S}/T_{\rm I}$ ratios. }
 \label{fig11}
\end{figure}

We compare the measured intensity of the CH$_3$OH transitions with those calculated by assuming the local thermodynamic equilibrium (LTE) approximation in order to check accuracy of the intensity calibration of the SUMIRE.   The integrated intensity of molecular emission ($I$) can be calculated under the optically thin and LTE conditions as: 
\begin{equation}
\label{eq8}
I = \frac{8 \pi^3 S \mu^2 \nu N}{3k_{\rm B} U(T)} \left(1-\frac{1-\exp \left(\frac{h \nu}{k_{\rm B}T}\right)}{1-\exp \left(\frac{h \nu}{k_{\rm B}T_{\rm cold}}\right)} \right) \exp\left(-\frac{E_{\rm u}}{k_{\rm B}T}\right),
\end{equation}
where $S$ is the line strength, $\mu$ is the dipole moment, $\nu$ is the frequency of the transition line, $N$ is the column density, $k_{\rm B}$ is the Boltzmann constant, $T$ is the gas temperature, $U(T)$ is the partition function at the temperature of $T$, $h$ is the Planck constant, $T_{\rm cold}$ is the temperature of cold load (77~K), and $E_{\rm u}$ is the upper state energy of the transition line.  We can calculate the peak intensity as 
\begin{equation}
\label{eq9}
T_{\rm mol} = \sqrt{\frac{4\ln 2}{\pi}} \frac{I}{\Delta v},
\end{equation}
where $\Delta v$ is the FWHM of the linewidth in the velocity unit. 

Figure~\ref{fig05} (b) shows the calculated spectrum.  Here, we assume the gas temperature of 296~K, the cold load temperature of 77~K, the CH$_3$OH column density of $2.3 \times 10^{16}$~cm$^{-2}$, and a line shape of the Gaussian profile with the $\Delta v$ of 0.7~km~s$^{-1}$.  The column density is evaluated from the gas pressure of 0.48~Pa, the gas temperature of 296~K, and the length of the sample cell of 200~cm.  We use the frequencies, the upper state energies, and the line intensities listed in the CDMS and adopt the partition function value of 36831.557 at 296~K given by \citet{Brauer2012}, because their partition function involves contributions from vibration modes.  As compared with the calculated spectrum in Figure~\ref{fig05}(b), our measured CH$_3$OH spectral lines shown in Figure~\ref{fig05}(a) are reproduced quite well. 

Figure~\ref{fig09} shows a correlation plot between the measured and calculated integrated intensities.  The measured integrated intensities are almost consistent with those of the calculated ones, although some values deviate.  

Possible origins of the deviations are the baseline ripples originated by the standing wave, a noise contamination by side lobes of the measurement beam (spillover effect), and variation of the image rejection ratio of the receiver.  Among them, the side lobe does not affect the measured intensities.  If we assume the main beam efficiency of $\eta$, the intensity of the hot-load ($S^{\prime}_{\rm hot}$), the cold-load ($S^{\prime}_{\rm cold}$), and the molecular emission ($S^{\prime}_{mol}$) are modified as:
\begin{eqnarray}
S^{\prime}_{\rm cold} & = & k (\eta T_{\rm cold} + (1-\eta) T_{\rm amb} + T_{\rm sys}) \\
S^{\prime}_{\rm hot} & = & k (\eta T_{\rm hot} + (1-\eta) T_{\rm amb} + T_{\rm sys}) \\
S^{\prime}_{\rm mol} & = & k (\eta (T_{\rm mol} + T_{\rm cold}) + (1-\eta) T_{\rm amb} + T_{\rm sys}),  
\end{eqnarray}
where $T_{\rm amb}$ is the room temperature (296~K).  Here, the side lobes of the beam are terminated to the room temperature body.  As a result, we obtain $T_{\rm mol}$ by the intensity calibration (equation~\ref{eq4}) as  
\begin{equation}
\label{eq13}
\frac{S^{\prime}_{\rm mol} - S^{\prime}_{\rm cold}}{S^{\prime}_{\rm hot} - S^{\prime}_{\rm cold}} \times (T_{\rm hot} - T_{\rm cold}) = T_{\rm mol}.
\end{equation}
Indeed, the term $(1-\eta)T_{\rm amb}$ and the main beam efficiency $\eta$, which are the contributions of the side lobe, are canceled out.

Another possibility is variation of the rejection ratio (IRR) of the 2SB receiver.  Because the 2SB receiver cannot perfectly separate the USB and LSB signals, the signal in one sideband (image sideband) is contaminated to the other sideband (signal sideband).  As the result, the intensity in the signal sideband is weakened by the leakage.  This effect depends on the frequency and the receiver tuning.  To inspect the IRR, we evaluate the intensity ratios of an emission line in the signal sideband ($T_{\rm S}$) to a corresponding ghost line in the image sideband ($T_{\rm I}$) by using strong CH$_3$OH lines ($T_{\rm peak}>1.5$~K), since the $T_{\rm S}/T_{\rm I}$ ratio is appropriate indicator of the IRR.  We obtain $T_{\rm S}$ and $T_{\rm I}$ by the Gaussian fitting and calculate the $T_{\rm S}/T_{\rm I}$ ratios.  Figure~\ref{fig10} shows the $T_{\rm S}/T_{\rm I}$ ratio as a function of the signal sideband frequency.  Most of the $T_{\rm S}/T_{\rm I}$ ratio is higher than 10~dB, indicating that the leakage is expected to be less than 10~\%.  Figure~\ref{fig11} is the ratios of the experimental integrated intensities to the LTE values as a function of the $T_{\rm S}/T_{\rm I}$ ratio.  Since the ratios tend to be low for the $T_{\rm S}/T_{\rm I}$ ratios lower than $\sim 10$~dB, the intensity deviation from the calculated value can be explained partly by this effect.  For a more precise measurement of the intensities, we need to correct the IRR.

\subsection{Rotation Diagram of CH$_3$OH}
We derive the rotation temperature and the column density of CH$_3$OH in the sample cell by using the rotation diagram method.  Under conditions of $\frac{h\nu}{k_{\rm B}T} \ll 1$ and $\frac{h\nu}{k_{\rm B}T_{\rm cold}} \ll 1$, Equation~(\ref{eq8}) can be approximated to be, 
\begin{equation}
\label{eq14}
I = \frac{8 \pi^3 S \mu^2 \nu N}{3k_{\rm B} U(T)} \left(1-\frac{T_{\rm cold}}{T} \right) \exp\left(-\frac{E_{\rm u}}{k_{\rm B}T}\right).
\end{equation}
Therefore, equation~(\ref{eq14}) is transformed as
\begin{equation}
\label{eq15}
\ln \left( \frac{3k_{\rm B}I}{8 \pi^3 S \mu^2 \nu} \right) = \ln \left( \frac{N}{U(T)} \left(1-\frac{T_{\rm cold}}{T}\right)\right) - \frac{E_{\rm u}}{k_{\rm B}T}.
\end{equation}
Here, the term $1-T_{\rm cold}/T$ usually approximates to 1 for the analysis of interstellar molecular clouds whose background temperature is that of the cosmic microwave background of $\sim 2.7$~K.  However, this approximation is not valid in this experiment because of higher background temperature of 77~K.  

\begin{figure}[t]
 \begin{center}
  \includegraphics[width=6.5cm]{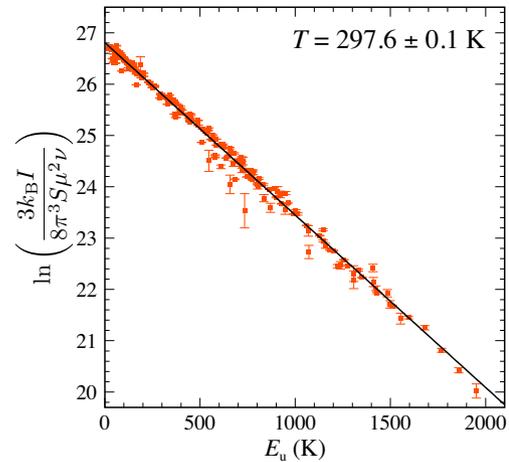}
 \end{center}
 \caption{Rotation diagram of CH$_3$OH. }
 \label{fig12}
\end{figure}

Figure~\ref{fig12} shows the rotation diagram of CH$_3$OH.  In this diagram, we exclude the data for which more than two transition lines are blended.  $S\mu^2$ and $E_{\rm u}$ are taken from the CDMS.  Since the transition lines with $v_t=3$ are not included in the CDMS, we also exclude them from the rotation diagram analysis.  The rotation temperature and the column density are determined to be $297.6 \pm 0.1$~K and $(2.2 \pm 0.2) \times 10^{16}$~cm$^{-2}$, respectively.  Here, we employ the partition function of 37254.1 at 297.6~K by a linear interpolation with the partition functions at fixed temperatures of 296, 297, and 300~K given by \citet{Brauer2012}.

The obtained rotation temperature is almost the same as the cell temperature of $296-298$~K during the experiments.  This result indicates that the relative intensity of spectra measured with the SUMIRE is reasonably precise.  The column density is also consistent with the column density of $2.3 \times 10^{16}$~cm$^{-2}$ evaluated from the gas pressure and the temperature within the $1\sigma$ uncertainty.  This is natural consequence in the view that the measured spectrum can be well reproduced by the calculation (Figure~\ref{fig05}).  

\begin{figure}[t]
 \begin{center}
  \includegraphics[width=6.5cm]{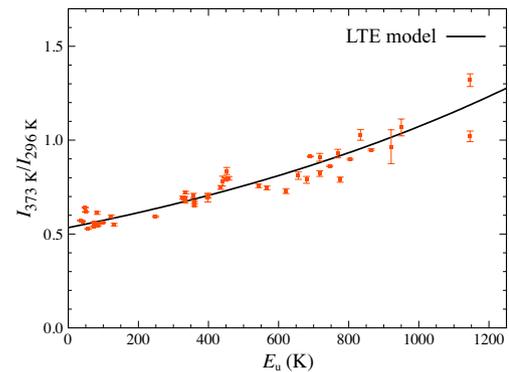}
 \end{center}
 \caption{The ratios of integrated intensity of CH$_3$OH between at the gas temperature of 373~K and 296~K as a function of $E_{\rm u}$.  The solid line is the ratio calculated by the LTE model.}
 \label{fig13}
\end{figure}

\subsection{Temperature Dependences}
As another test of the intensity calibration, the CH$_3$OH spectrum was measured at the sample pressure of 0.4-0.5~Pa and the gas temperature of 373~K by using the tape heaters wound around the cell.  In this experiment, the frequency range from 240~GHz to 244~GHz is covered with the two 2500~MHz XFFTSs with the 1st LO frequency of 248~GHz.  The total integration time is 30~minutes, consisting of 6 times the 5~minutes sample measurements.  Figure~\ref{fig13} shows the ratios of integrated intensities between at the gas temperature of 373~K ($I_{\rm 373\,K}$) and 296~K ($I_{\rm 296\,K}$) as a function of the $E_{\rm u}$. The ratios are found to increase with the $E_{\rm u}$.  This trend can be reproduced by the LTE model (solid line in Figure~\ref{fig13}) calculated by using the equation (\ref{eq8}).  This result verifies that the intensity accuracy of our system is enough to detect the gas temperature difference of $\sim 70$~K.  Moreover, this indicates that upper state energies of unidentified emission lines detected by our system can be estimated by measuring a line ratio between two different gas temperatures.

\section{Summary}
We have developed the emission-type millimeter and submillimeter spectrometer using an SIS mixer receiver, SUMIRE, for laboratory measurements of molecular spectral lines in the 210 - 270 GHz band.  The ALMA-type cartridge receiver is employed in this spectrometer so that the frequency band can readily be changed by exchanging the receiver cartridge.  The two types of XFFTS with the bandwidth of 2.5~GHz and 0.5~GHz are used as a backend for the broadband measurement and high frequency-resolution measurements, respectively.  The frequency accuracy is guaranteed by the rubidium clock assisted with the GPS system.  The absolute intensity scale is calibrated by using the blackbody radiations at the room temperature and the liquid nitrogen temperature.  Test measurements of SUMIRE have been conducted for CH$_3$CN, HDO, and CH$_3$OH to reveal its basic performance. 

In the CH$_3$CN measurements, we confirm that the frequencies observed with SUMIRE are consistent with those listed in the CDMS within 1~kHz for intense lines.  We also confirm that the frequencies measured with the 2500~MHz XFFTS are consistent with those with the 500~MHz XFFTS. 

We have measured the transition frequencies of five HDO lines and two HD$^{18}$O lines accurately by using 500~MHz XFFTS.  They are found to significantly deviate from the frequencies listed in the database by 18-145~kHz. These new frequencies are useful for astronomical observations and for refining the molecular constants of these species. 

We have conducted the spectral scan measurement from 216 to 264~GHz by using the 2500~MHz XFFTS for CH$_3$OH.  We have identified 242 transitions of CH$_3$OH including 16 transitions of the third torsional state from 295 emission lines.  The measured frequencies are compared with those reported in the CDMS.  Although a reasonable agreement is obtained for the strong lines, frequencies of high excitation lines tend to deviate from those listed in the CDMS.  This result clearly demonstrates importance of the direct measurement of transition frequencies.  Intensities of the measured spectral lines are consistent with those expected from the column density of CH$_3$OH and the gas temperature.  The measurement has also been done at an elevated temperature (373~K), and the temperature dependence of line intensities is confirmed to be consistent with the expectation from the upper state energies of the lines.

\newpage


\begin{ack}
We are very grateful to Satoshi~Yamamoto for extensive helps, suggestions, and discussions of this experiment.  We also very grateful to Tatsuhiko~Satoh and Masa~Takegahara for great helps for developing the SUMIRE system.  We would like to acknowledge Yoshinori~Uzawa for providing the SIS mixer tips and mixer block.  We thank Yuji~Ebisawa, Hidetoshi~Katori, Makoto~Nagai, Yuri~Nishimura, Yoko~Oya, Osamu~Oguchi, Tastuya~Soma, and Kento~Yoshida for great helps.  We would also like to thank Stephan Schlemmer for fruitful discussions.  This study is supported by a Grant-in-Aid from the Ministry of Education, Culture, Sports, Science, and Technology of Japan (No. 25108005, and 18H05222).  Y.W. is supported by Tsukuba Basic Research Support Program Type A from University of Tsukuba.  
\end{ack}


\appendix 
\section*{Frequency List of CH$_3$OH}
We summarise the results of the Gaussian fittings to the CH$_3$OH spectrum in Table~\ref{tab06}.  The catalogue values are quoted from the CDMS \citep{cdms2005} and \citet{Pearson2009} for CH$_3$OH ($v_t$ = 0, 1, and 2) and $^{13}$CH$_3$OH, and CH$_3$OH ($v_t$ = 3), respectively. 
{\small
\begin{longtable}{llrllll}
\caption{Line list of CH$_3$OH measurement with the SUMIRE.}\label{tab06}
\hline              
  Freq. Obs. $^{\rm a}$ & Freq. Cat. $^{\rm a,b}$ & Obs.-Cat.$^{\rm c}$ & $\sigma$ $^{\rm a,d}$ & $T$ $^{\rm a,e}$ & Mol. $^{\rm f}$ & Transition \\ 
  (MHz)      & (MHz)       & (MHz)     & (km s$^{-1}$) & (K) &  & \\ 
\endfirsthead
\hline
  Freq. Obs. $^{\rm a}$ & Freq. Cat. $^{\rm a,b}$ & Obs.-Cat.$^{\rm c}$ & $\sigma$ $^{\rm a,d}$ & $T$ $^{\rm a,e}$ & Mol. $^{\rm f}$ & Transition \\ 
\hline
\endhead
\hline
\endfoot
\hline
\multicolumn{7}{l}{{\small $^{\rm a}$ The numbers in parentheses represent $1\sigma$ error.}} \\
\multicolumn{7}{l}{{\small $^{\rm b}$ The frequency given by the catalogues.}} \\
\multicolumn{7}{l}{{\small $^{\rm c}$ Difference of frequency between the observation and the catalogue.}} \\
\multicolumn{7}{l}{{\small $^{\rm d}$ The standard deviation of the Gaussian function derived by the Gaussian fitting.}} \\
\multicolumn{7}{l}{{\small $^{\rm e}$ Peak intensity derived by the Gaussian fitting.}} \\
\multicolumn{7}{l}{{\small $^{\rm f}$ UI denotes unidentified line.}} \\
\multicolumn{7}{l}{{\small $^{\rm g}$ The line is blended with two transitions.}} \\
\endlastfoot
 \hline
216133.354 (0.024) & 216131.834 (0.155) &  1.520 & 0.321 (0.035) &  0.12 (0.01) & CH$_3$OH & $35_{-6}-34_{-7} \, {\rm E} \, v_t = 0$\\ 
216945.577 (0.002) & 216945.521 (0.012) &  0.056 & 0.335 (0.003) &  1.96 (0.01) & CH$_3$OH & $5_{1}-4_{2} \, {\rm E} \, v_t = 0$\\ 
216952.478 (0.021) &  &  & 0.328 (0.030) &  0.13 (0.01) & UI &  \\ 
217044.615 (0.050) & 217044.616 (0.064) & -0.001 & 0.357 (0.072) &  0.06 (0.01) & $^{13}$CH$_3$OH & $14_{1}-13_{2} \, {\rm A^-} \, v_t = 0$\\ 
217299.209 (0.002) & 217299.205 (0.017) &  0.004 & 0.319 (0.002) &  3.13 (0.02) & CH$_3$OH & $6_{1}-7_{2} \, {\rm A^-} \, v_t = 1$\\ 
217642.734 (0.002) $^{\rm g}$ & 217642.677 (0.022) &  0.057 & 0.323 (0.003) &  1.72 (0.01) & CH$_3$OH & $15_{6}-16_{5} \, {\rm A^-} \, v_t = 1$\\ 
217642.734 (0.002) $^{\rm g}$ & 217642.678 (0.022) &  0.056 & 0.323 (0.003) &  1.72 (0.01) & CH$_3$OH & $15_{6}-16_{5} \, {\rm A^+} \, v_t = 1$\\ 
217886.524 (0.002) & 217886.504 (0.022) &  0.020 & 0.327 (0.003) &  4.33 (0.03) & CH$_3$OH & $20_{1}-20_{0} \, {\rm E} \, v_t = 0$\\ 
218440.045 (0.002) & 218440.063 (0.013) & -0.018 & 0.332 (0.003) &  6.04 (0.04) & CH$_3$OH & $4_{2}-3_{1} \, {\rm E} \, v_t = 0$\\ 
219983.696 (0.002) & 219983.675 (0.017) &  0.021 & 0.332 (0.003) &  1.39 (0.01) & CH$_3$OH & $25_{3}-24_{4} \, {\rm E} \, v_t = 0$\\ 
219993.657 (0.002) & 219993.658 (0.018) & -0.001 & 0.326 (0.003) &  1.23 (0.01) & CH$_3$OH & $23_{5}-22_{6} \, {\rm E} \, v_t = 0$\\ 
220078.521 (0.002) & 220078.561 (0.008) & -0.040 & 0.329 (0.002) &  6.55 (0.04) & CH$_3$OH & $8_{0}-7_{1} \, {\rm E} \, v_t = 0$\\ 
220401.367 (0.002) & 220401.317 (0.014) &  0.050 & 0.333 (0.003) &  2.13 (0.02) & CH$_3$OH & $10_{-5}-11_{-4} \, {\rm E} \, v_t = 0$\\ 
220886.591 (0.002) & 220886.784 (0.080) & -0.193 & 0.320 (0.003) &  1.87 (0.02) & CH$_3$OH & $31_{-2}-31_{1} \, {\rm E} \, v_t = 0$\\ 
221631.803 (0.023) &  &  & 0.335 (0.033) &  0.17 (0.01) & UI &  \\ 
222467.943 (0.022) & 222468.344 (0.718) & -0.401 & 0.184 (0.030) &  0.12 (0.02) & $^{13}$CH$_3$OH & $21_{1}-21_{0} \, {\rm E} \, v_t = 0$\\ 
222722.850 (0.002) & 222722.856 (0.029) & -0.006 & 0.321 (0.002) &  3.49 (0.02) & CH$_3$OH & $16_{2}-15_{1} \, {\rm A^+} \, v_t = 1$\\ 
222829.038 (0.030) $^{\rm g}$ & 222828.860 (0.044) &  0.178 & 0.293 (0.041) &  0.10 (0.01) & CH$_3$OH & $17_{8}-18_{7} \, {\rm A^+} \, v_t = 1$\\ 
222829.038 (0.030) $^{\rm g}$ & 222828.860 (0.044) &  0.178 & 0.293 (0.041) &  0.10 (0.01) & CH$_3$OH & $17_{8}-18_{7} \, {\rm A^-} \, v_t = 1$\\ 
223507.647 (0.023) & 223507.831 (0.060) & -0.184 & 0.372 (0.032) &  0.25 (0.02) & CH$_3$OH & $31_{-7}-30_{-8} \, {\rm E} \, v_t = 0$\\ 
224001.320 (0.005) & 224001.371 (0.032) & -0.051 & 0.311 (0.004) &  0.91 (0.02) & CH$_3$OH & $21_{4}-22_{5} \, {\rm A^-} \, v_t = 1$\\ 
224001.920 (0.005) & 224001.970 (0.032) & -0.050 & 0.311 (0.004) &  0.93 (0.02) & CH$_3$OH & $21_{4}-22_{5} \, {\rm A^+} \, v_t = 1$\\ 
224699.448 (0.002) & 224699.408 (0.016) &  0.040 & 0.327 (0.003) &  3.13 (0.02) & CH$_3$OH & $20_{-2}-19_{-3} \, {\rm E} \, v_t = 0$\\ 
224843.312 (0.005) & 224843.232 (0.021) &  0.080 & 0.324 (0.007) &  0.72 (0.01) & CH$_3$OH & $25_{7}-26_{6} \, {\rm E} \, v_t = 0$\\ 
225313.445 (0.006) & 225313.395 (0.024) &  0.050 & 0.317 (0.008) &  0.71 (0.02) & CH$_3$OH & $25_{-7}-26_{-6} \, {\rm E} \, v_t = 0$\\ 
226820.378 (0.016) &  &  & 0.302 (0.022) &  0.18 (0.01) & UI &  \\ 
227094.745 (0.002) & 227094.747 (0.023) & -0.002 & 0.323 (0.002) &  4.91 (0.03) & CH$_3$OH & $21_{1}-21_{0} \, {\rm E} \, v_t = 0$\\ 
227527.517 (0.018) & 227528.092 (0.080) & -0.575 & 0.297 (0.025) &  0.20 (0.01) & CH$_3$OH & $31_{8}-32_{7} \, {\rm E} \, v_t = 0$\\ 
227814.405 (0.002) & 227814.528 (0.017) & -0.123 & 0.325 (0.002) &  4.81 (0.03) & CH$_3$OH & $16_{1}-15_{2} \, {\rm A^+} \, v_t = 0$\\ 
228257.185 (0.024) & 228256.537 (0.117) &  0.648 & 0.322 (0.033) &  0.09 (0.01) & CH$_3$OH & $31_{-4}-32_{-3} \, {\rm E} \, v_t = 1$\\ 
228611.600 (0.008) & 228611.615 (0.034) & -0.015 & 0.300 (0.011) &  0.25 (0.01) & CH$_3$OH & $26_{9}-27_{8} \, {\rm E} \, v_t = 0$\\ 
229356.822 (0.023) & 229356.599 (0.035) &  0.223 & 0.298 (0.032) &  0.08 (0.01) & CH$_3$OH & $20_{2}-19_{3} \, {\rm E} \, v_t = 2$\\ 
229589.078 (0.002) & 229589.056 (0.012) &  0.022 & 0.324 (0.002) &  2.95 (0.02) & CH$_3$OH & $15_{4}-16_{3} \, {\rm E} \, v_t = 0$\\ 
229758.775 (0.002) & 229758.756 (0.012) &  0.019 & 0.328 (0.002) &  7.97 (0.05) & CH$_3$OH & $8_{-1}-7_{0} \, {\rm E} \, v_t = 0$\\ 
229864.164 (0.002) & 229864.121 (0.014) &  0.043 & 0.323 (0.003) &  1.71 (0.01) & CH$_3$OH & $19_{5}-20_{4} \, {\rm A^+} \, v_t = 0$\\ 
229939.141 (0.002) & 229939.095 (0.014) &  0.046 & 0.323 (0.003) &  1.76 (0.01) & CH$_3$OH & $19_{5}-20_{4} \, {\rm A^-} \, v_t = 0$\\ 
230027.033 (0.002) & 230027.047 (0.011) & -0.014 & 0.327 (0.003) &  1.46 (0.01) & CH$_3$OH & $3_{-2}-4_{-1} \, {\rm E} \, v_t = 0$\\ 
230292.298 (0.011) & 230292.196 (0.018) &  0.102 & 0.354 (0.015) &  0.24 (0.01) & CH$_3$OH & $22_{2}-21_{-3} \, {\rm E} \, v_t = 0$\\ 
230368.701 (0.002) & 230368.763 (0.018) & -0.062 & 0.323 (0.003) &  1.30 (0.01) & CH$_3$OH & $22_{4}-21_{5} \, {\rm E} \, v_t = 0$\\ 
230817.587 (0.034) & 230817.637 (0.053) & -0.050 & 0.289 (0.046) &  0.07 (0.01) & CH$_3$OH & $18_{6}-18_{7} \, {\rm E} \, v_t = 1$\\ 
231006.622 (0.044) $^{\rm g}$ & 231005.758 (0.282) &  0.864 & 0.418 (0.061) &  0.06 (0.01) & CH$_3$OH & $38_{7}-37_{8} \, {\rm A^-} \, v_t = 0$\\ 
231006.622 (0.044) $^{\rm g}$ & 231006.232 (0.282) &  0.390 & 0.418 (0.061) &  0.06 (0.01) & CH$_3$OH & $38_{7}-37_{8} \, {\rm A^+} \, v_t = 0$\\ 
231281.133 (0.002) & 231281.110 (0.012) &  0.023 & 0.327 (0.002) &  3.27 (0.02) & CH$_3$OH & $10_{2}-9_{3} \, {\rm A^-} \, v_t = 0$\\ 
231580.646 (0.032) & 231580.650 (0.048) & -0.004 & 0.434 (0.045) &  0.09 (0.01) & CH$_3$OH & $17_{6}-17_{7} \, {\rm E} \, v_t = 1$\\ 
232180.360 (0.019) & 232180.378 (0.043) & -0.018 & 0.264 (0.026) &  0.20 (0.02) & CH$_3$OH & $16_{6}-16_{7} \, {\rm E} \, v_t = 1$\\ 
232418.557 (0.002) & 232418.521 (0.012) &  0.036 & 0.327 (0.003) &  4.16 (0.03) & CH$_3$OH & $10_{2}-9_{3} \, {\rm A^+} \, v_t = 0$\\ 
232422.934 (0.028) & 232422.803 (0.046) &  0.131 & 0.337 (0.038) &  0.18 (0.02) & CH$_3$OH & $7_{6}-7_{7} \, {\rm E} \, v_t = 1$\\ 
232538.383 (0.035) &  &  & 0.456 (0.048) &  0.16 (0.01) & UI &  \\ 
232624.844 (0.018) & 232624.811 (0.038) &  0.033 & 0.274 (0.024) &  0.21 (0.01) & CH$_3$OH & $15_{6}-15_{7} \, {\rm E} \, v_t = 1$\\ 
232645.275 (0.018) & 232645.103 (0.042) &  0.172 & 0.327 (0.024) &  0.26 (0.02) & CH$_3$OH & $8_{6}-8_{7} \, {\rm E} \, v_t = 1$\\ 
232783.528 (0.002) & 232783.446 (0.012) &  0.082 & 0.323 (0.002) &  3.48 (0.02) & CH$_3$OH & $18_{3}-17_{4} \, {\rm A^+} \, v_t = 0$\\ 
232847.246 (0.009) & 232847.103 (0.038) &  0.143 & 0.315 (0.012) &  0.37 (0.01) & CH$_3$OH & $9_{6}-9_{7} \, {\rm E} \, v_t = 1$\\ 
232925.490 (0.015) & 232925.435 (0.034) &  0.055 & 0.323 (0.020) &  0.24 (0.01) & CH$_3$OH & $14_{6}-14_{7} \, {\rm E} \, v_t = 1$\\ 
232945.831 (0.002) & 232945.797 (0.012) &  0.034 & 0.328 (0.002) &  4.21 (0.03) & CH$_3$OH & $10_{-3}-11_{-2} \, {\rm E} \, v_t = 0$\\ 
233012.015 (0.011) & 233011.878 (0.035) &  0.137 & 0.330 (0.015) &  0.35 (0.01) & CH$_3$OH & $10_{6}-10_{7} \, {\rm E} \, v_t = 1$\\ 
233096.843 (0.011) & 233096.805 (0.032) &  0.038 & 0.290 (0.014) &  0.31 (0.01) & CH$_3$OH & $13_{6}-13_{7} \, {\rm E} \, v_t = 1$\\ 
233121.272 (0.009) & 233121.162 (0.032) &  0.110 & 0.319 (0.012) &  0.36 (0.01) & CH$_3$OH & $11_{6}-11_{7} \, {\rm E} \, v_t = 1$\\ 
233123.892 (0.030) &  &  & 0.324 (0.012) &  0.11 (0.01) & UI &  \\ 
233155.967 (0.008) & 233155.874 (0.031) &  0.093 & 0.309 (0.011) &  0.36 (0.01) & CH$_3$OH & $12_{6}-12_{7} \, {\rm E} \, v_t = 1$\\ 
233795.748 (0.002) & 233795.666 (0.012) &  0.082 & 0.324 (0.003) &  3.40 (0.02) & CH$_3$OH & $18_{3}-17_{4} \, {\rm A^-} \, v_t = 0$\\ 
233916.984 (0.005) $^{\rm g}$ & 233916.950 (0.022) &  0.034 & 0.324 (0.006) &  0.81 (0.01) & CH$_3$OH & $13_{3}-14_{4} \, {\rm A^-} \, v_t = 2$\\ 
233916.984 (0.005) $^{\rm g}$ & 233917.018 (0.022) & -0.034 & 0.324 (0.007) &  0.81 (0.01) & CH$_3$OH & $13_{3}-14_{4} \, {\rm A^+} \, v_t = 2$\\ 
234011.612 (0.026) & 234011.580 (0.050) &  0.032 & 0.311 (0.034) &  0.12 (0.01) & $^{13}$CH$_3$OH & $5_{1}-4_{1} \, {\rm A^+} \, v_t = 0$\\ 
234523.139 (0.003) & 234523.365 (0.107) & -0.226 & 0.318 (0.004) &  1.63 (0.01) & CH$_3$OH & $32_{-2}-32_{1} \, {\rm E} \, v_t = 0$\\ 
234683.396 (0.002) & 234683.370 (0.012) &  0.026 & 0.323 (0.003) &  2.42 (0.02) & CH$_3$OH & $4_{2}-5_{1} \, {\rm A^-} \, v_t = 0$\\ 
234698.478 (0.005) & 234698.519 (0.015) & -0.041 & 0.334 (0.007) &  0.79 (0.01) & CH$_3$OH & $5_{-4}-6_{-3} \, {\rm E} \, v_t = 0$\\ 
235881.237 (0.038) & 235881.170 (0.050) &  0.067 & 0.302 (0.051) &  0.11 (0.01) & $^{13}$CH$_3$OH & $5_{0}-4_{0} \, {\rm E} \, v_t = 0$\\ 
235938.195 (0.023) & 235938.220 (0.050) & -0.025 & 0.271 (0.030) &  0.15 (0.01) & $^{13}$CH$_3$OH & $5_{-1}-4_{-1} \, {\rm E} \, v_t = 0$\\ 
235960.330 (0.029) & 235960.370 (0.050) & -0.040 & 0.265 (0.038) &  0.12 (0.01) & $^{13}$CH$_3$OH & $5_{0}-4_{0} \, {\rm A^+} \, v_t = 0$\\ 
236006.133 (0.053) & 236006.100 (0.050) &  0.033 & 0.225 (0.036) &  0.06 (0.01) & $^{13}$CH$_3$OH & $5_{3}-4_{3} \, {\rm E} \, v_t = 0$\\ 
236008.433 (0.032) & 236008.390 (0.050) &  0.043 & 0.225 (0.036) &  0.10 (0.01) & $^{13}$CH$_3$OH & $5_{2}-4_{2} \, {\rm A^-} \, v_t = 0$\\ 
236016.542 (0.038) & 236016.550 (0.050) & -0.008 & 0.210 (0.050) &  0.08 (0.02) & $^{13}$CH$_3$OH & $5_{-3}-4_{-3} \, {\rm E} \, v_t = 0$\\ 
236041.408 (0.021) & 236041.400 (0.050) &  0.008 & 0.253 (0.027) &  0.14 (0.01) & $^{13}$CH$_3$OH & $5_{1}-4_{1} \, {\rm E} \, v_t = 0$\\ 
236049.604 (0.035) & 236049.520 (0.050) &  0.084 & 0.316 (0.047) &  0.09 (0.01) & $^{13}$CH$_3$OH & $5_{2}-4_{2} \, {\rm A^+} \, v_t = 0$\\ 
236062.067 (0.031) & 236062.000 (0.050) &  0.067 & 0.297 (0.024) &  0.11 (0.01) & $^{13}$CH$_3$OH & $5_{-2}-4_{-2} \, {\rm E} \, v_t = 0$\\ 
236062.905 (0.030) & 236062.850 (0.050) &  0.055 & 0.297 (0.024) &  0.12 (0.01) & $^{13}$CH$_3$OH & $5_{2}-4_{2} \, {\rm E} \, v_t = 0$\\ 
236353.104 (0.014) & 236353.064 (0.034) &  0.040 & 0.256 (0.019) &  0.23 (0.01) & $^{13}$CH$_3$OH & $11_{0}-10_{1} \, {\rm A^+} \, v_t = 0$\\ 
236648.865 (0.025) & 236648.729 (0.028) &  0.136 & 0.146 (0.033) &  0.10 (0.02) & CH$_3$OH & $11_{7}-10_{8} \, {\rm E} \, v_t = 1$\\ 
236853.416 (0.003) & 236853.355 (0.028) &  0.061 & 0.313 (0.003) &  2.02 (0.02) & CH$_3$OH & $23_{-3}-22_{-2} \, {\rm E} \, v_t = 1$\\ 
236936.131 (0.002) & 236936.089 (0.015) &  0.042 & 0.322 (0.002) &  7.20 (0.04) & CH$_3$OH & $14_{1}-13_{2} \, {\rm A^-} \, v_t = 0$\\ 
237129.378 (0.001) & 237129.402 (0.023) & -0.024 & 0.322 (0.002) &  4.63 (0.02) & CH$_3$OH & $22_{1}-22_{0} \, {\rm E} \, v_t = 0$\\ 
237970.412 (0.003) & 237970.421 (0.036) & -0.009 & 0.322 (0.003) &  1.38 (0.01) & CH$_3$OH & $17_{-5}-18_{-6} \, {\rm E} \, v_t = 1$\\ 
237983.351 (0.029) $^{\rm g}$ & 237983.380 (0.050) & -0.029 & 0.343 (0.038) &  0.10 (0.01) & $^{13}$CH$_3$OH & $5_{1}-4_{1} \, {\rm A^-} \, v_t = 0$\\ 
237983.351 (0.030) $^{\rm g}$ & 237983.380 (0.050) & -0.029 & 0.338 (0.039) &  0.10 (0.01) & $^{13}$CH$_3$OH & $5_{1}-4_{1} \, {\rm A^-} \, v_t = 0$\\ 
238177.559 (0.028) &  &  & 0.281 (0.037) &  0.12 (0.01) & UI &  \\ 
238440.922 (0.003) $^{\rm g}$ & 238440.805 (0.037) &  0.117 & 0.335 (0.004) &  1.57 (0.02) & CH$_3$OH & $25_{5}-24_{6} \, {\rm A^+} \, v_t = 1$\\ 
238440.922 (0.003) $^{\rm g}$ & 238440.974 (0.037) & -0.052 & 0.335 (0.004) &  1.57 (0.01) & CH$_3$OH & $25_{5}-24_{6} \, {\rm A^-} \, v_t = 1$\\ 
238665.496 (0.002) & 238665.976 (0.046) & -0.480 & 0.321 (0.002) &  2.23 (0.01) & CH$_3$OH & $30_{3}-30_{2} \, {\rm A^{-+}} \, v_t = 0$\\ 
238728.758 (0.002) & 238729.425 (0.056) & -0.667 & 0.314 (0.003) &  1.76 (0.01) & CH$_3$OH & $31_{3}-31_{2} \, {\rm A^{-+}} \, v_t = 0$\\ 
238806.703 (0.020) &  &  & 0.343 (0.027) &  0.16 (0.01) & UI &  \\ 
238890.095 (0.002) & 238890.424 (0.039) & -0.329 & 0.316 (0.002) &  2.61 (0.01) & CH$_3$OH & $29_{3}-29_{2} \, {\rm A^{-+}} \, v_t = 0$\\ 
239058.229 (0.035) &  &  & 0.289 (0.046) &  0.08 (0.01) & UI &  \\ 
239141.863 (0.003) & 239142.749 (0.067) & -0.886 & 0.315 (0.004) &  1.38 (0.02) & CH$_3$OH & $32_{3}-32_{2} \, {\rm A^{-+}} \, v_t = 0$\\ 
239230.140 (0.043) &  &  & 0.250 (0.056) &  0.07 (0.01) & UI &  \\ 
239246.625 (0.042) &  &  & 0.287 (0.055) &  0.07 (0.01) & UI &  \\ 
239265.330 (0.037) &  &  & 0.321 (0.049) &  0.08 (0.01) & UI &  \\ 
239344.846 (0.002) & 239345.054 (0.033) & -0.208 & 0.314 (0.002) &  3.15 (0.02) & CH$_3$OH & $28_{3}-28_{2} \, {\rm A^{-+}} \, v_t = 0$\\ 
239397.954 (0.030) & 239397.997 (0.016) & -0.043 & 0.294 (0.040) &  0.10 (0.01) & CH$_3$OH & $16_{-3}-17_{0} \, {\rm E} \, v_t = 0$\\ 
239660.114 (0.027) & 239659.467 (0.118) &  0.647 & 0.324 (0.036) &  0.12 (0.01) & CH$_3$OH & $33_{-2}-33_{2} \, {\rm E} \, v_t = 0$\\ 
239731.302 (0.002) $^{\rm g}$ & 239731.362 (0.017) & -0.060 & 0.317 (0.002) &  2.84 (0.02) & CH$_3$OH & $16_{7}-17_{6} \, {\rm A^+} \, v_t = 0$\\ 
239731.302 (0.002) $^{\rm g}$ & 239731.363 (0.017) & -0.061 & 0.317 (0.002) &  2.84 (0.02) & CH$_3$OH & $16_{7}-17_{6} \, {\rm A^-} \, v_t = 0$\\ 
239746.226 (0.002) & 239746.219 (0.004) &  0.007 & 0.322 (0.002) &  9.69 (0.05) & CH$_3$OH & $5_{1}-4_{1} \, {\rm A^+} \, v_t = 0$\\ 
239970.224 (0.003) & 239971.367 (0.081) & -1.143 & 0.314 (0.004) &  1.21 (0.01) & CH$_3$OH & $33_{3}-33_{2} \, {\rm A^{-+}} \, v_t = 0$\\ 
239976.932 (0.002) & 239977.050 (0.028) & -0.117 & 0.316 (0.002) &  3.64 (0.02) & CH$_3$OH & $27_{3}-27_{2} \, {\rm A^{-+}} \, v_t = 0$\\ 
240219.789 (0.047) & 240219.837 & -0.048 & 0.394 (0.062) &  0.07 (0.01) & CH$_3$OH & $5_0-4_0\,{\rm A^+}\, v_t=3$ \\ 
240225.841 (0.038) &  &  & 0.320 (0.049) &  0.09 (0.01) & UI &  \\ 
240241.499 (0.002) & 240241.490 (0.014) &  0.009 & 0.331 (0.003) &  2.21 (0.02) & CH$_3$OH & $5_{3}-6_{2} \, {\rm E} \, v_t = 0$\\ 
240299.434 (0.031) & 240299.468 & -0.034 & 0.195 (0.039) &  0.09 (0.01) & CH$_3$OH & $5_{-1}-4_{-1}\,{\rm E}\, v_t=3$ \\ 
240321.142 (0.004) $^{\rm g}$ & 240321.199 (0.028) & -0.057 & 0.322 (0.005) &  0.78 (0.01) & CH$_3$OH & $27_{8}-28_{7} \, {\rm A^+} \, v_t = 0$\\ 
240321.142 (0.004) $^{\rm g}$ & 240321.205 (0.028) & -0.063 & 0.322 (0.005) &  0.78 (0.01) & CH$_3$OH & $27_{8}-28_{7} \, {\rm A^-} \, v_t = 0$\\ 
240348.487 (0.034) & 240348.439 &  0.048 & 0.297 (0.044) &  0.08 (0.01) & CH$_3$OH & $5_3-4_3\,{\rm E}\, v_t=3$ \\ 
240371.686 (0.032) & 240371.759 & -0.073 & 0.310 (0.041) &  0.10 (0.01) & CH$_3$OH & $5_2-4_2\,{\rm E}\, v_t=3$ \\ 
240433.390 (0.018) $^{\rm g}$ & 240433.384 &  0.006 & 0.267 (0.023) &  0.19 (0.01) & CH$_3$OH & $5_3-4_3\,{\rm A^+}\, v_t=3$ \\ 
240433.390 (0.018) $^{\rm g}$ & 240433.384 &  0.006 & 0.267 (0.023) &  0.19 (0.01) & CH$_3$OH & $5_3-4_3\,{\rm A^-}\, v_t=3$ \\ 
240437.298 (0.022) & 240437.300 & -0.002 & 0.290 (0.028) &  0.14 (0.01) & CH$_3$OH & $5_{-2}-4_{-2}\,{\rm E}\, v_t=3$ \\ 
240454.826 (0.004) & 240454.848 (0.010) & -0.022 & 0.309 (0.005) &  1.22 (0.02) & CH$_3$OH & $5_{1}-4_{1} \, {\rm A^+} \, v_t = 2$\\ 
240485.819 (0.042) & 240485.830 & -0.011 & 0.261 (0.054) &  0.06 (0.01) & CH$_3$OH & $5_{-4}-4_{-4}\,{\rm E}\, v_t=3$ \\ 
240515.827 (0.013) & 240515.796 &  0.031 & 0.303 (0.018) &  0.22 (0.01) & CH$_3$OH & $5_1-4_1\,{\rm A^-}\, v_t=3$ \\ 
240518.412 (0.013) & 240518.457 & -0.045 & 0.324 (0.018) &  0.24 (0.01) & CH$_3$OH & $5_1-4_1\,{\rm A^+}\, v_t=3$ \\ 
240588.096 (0.008) & 240588.074 &  0.022 & 0.299 (0.010) &  0.36 (0.01) & CH$_3$OH & $5_0-4_0\,{\rm E}\, v_t=3$ \\ 
240601.314 (0.026) & 240601.274 &  0.040 & 0.307 (0.034) &  0.12 (0.01) & CH$_3$OH & $5_4-4_4\,{\rm E}\, v_t=3$ \\ 
240649.899 (0.010) & 240649.928 & -0.029 & 0.293 (0.013) &  0.41 (0.01) & CH$_3$OH & $5_1-4_1\,{\rm E}\, v_t=3$ \\ 
240678.991 (0.009) & 240679.026 & -0.035 & 0.305 (0.011) &  0.34 (0.01) & CH$_3$OH & $5_{-3}-4_{-3}\,{\rm E}\, v_t=3$ \\ 
240700.823 (0.008) & 240700.850 & -0.027 & 0.301 (0.006) &  0.48 (0.01) & CH$_3$OH & $5_2-4_2\,{\rm A^+}\, v_t=3$ \\ 
240701.542 (0.008) & 240701.594 & -0.052 & 0.301 (0.006) &  0.47 (0.01) & CH$_3$OH & $5_2-4_2\,{\rm A^-}\, v_t=3$ \\ 
240738.878 (0.002) & 240738.926 (0.025) & -0.048 & 0.320 (0.002) &  4.55 (0.03) & CH$_3$OH & $26_{3}-26_{2} \, {\rm A^{-+}} \, v_t = 0$\\ 
240752.548 (0.006) & 240752.863 (0.030) & -0.315 & 0.307 (0.008) &  0.45 (0.01) & CH$_3$OH & $5_{-3}-4_{-3} \, {\rm E} \, v_t = 2$\\ 
240757.929 (0.003) $^{\rm g}$ & 240757.889 (0.013) &  0.040 & 0.320 (0.004) &  1.26 (0.01) & CH$_3$OH & $5_{2}-4_{2} \, {\rm A^+} \, v_t = 2$\\ 
240757.929 (0.003) $^{\rm g}$ & 240757.920 (0.013) &  0.009 & 0.320 (0.004) &  1.26 (0.01) & CH$_3$OH & $5_{2}-4_{2} \, {\rm A^-} \, v_t = 2$\\ 
240784.258 (0.012) & 240784.498 (0.015) & -0.240 & 0.256 (0.015) &  0.26 (0.01) & CH$_3$OH & $5_{4}-4_{4} \, {\rm E} \, v_t = 2$\\ 
240817.974 (0.004) & 240817.972 (0.008) &  0.002 & 0.311 (0.006) &  0.78 (0.01) & CH$_3$OH & $5_{1}-4_{1} \, {\rm E} \, v_t = 2$\\ 
240861.416 (0.013) & 240861.406 (0.006) &  0.010 & 0.312 (0.017) &  0.33 (0.01) & CH$_3$OH & $5_{-4}-4_{-4} \, {\rm E} \, v_t = 2$\\ 
240869.533 (0.004) & 240869.551 (0.008) & -0.018 & 0.305 (0.005) &  1.03 (0.01) & CH$_3$OH & $5_{0}-4_{0} \, {\rm E} \, v_t = 2$\\ 
240916.156 (0.003) $^{\rm g}$ & 240916.172 (0.006) & -0.016 & 0.311 (0.003) &  1.70 (0.02) & CH$_3$OH & $5_{3}-4_{3} \, {\rm A^-} \, v_t = 2$\\ 
240916.156 (0.003) $^{\rm g}$ & 240916.173 (0.006) & -0.017 & 0.311 (0.003) &  1.70 (0.02) & CH$_3$OH & $5_{3}-4_{3} \, {\rm A^+} \, v_t = 2$\\ 
240932.047 (0.004) $^{\rm g}$ & 240932.051 (0.006) & -0.004 & 0.309 (0.005) &  1.13 (0.01) & CH$_3$OH & $5_{4}-4_{4} \, {\rm A^+} \, v_t = 2$\\ 
240932.047 (0.004) $^{\rm g}$ & 240932.051 (0.006) & -0.004 & 0.309 (0.005) &  1.13 (0.01) & CH$_3$OH & $5_{4}-4_{4} \, {\rm A^-} \, v_t = 2$\\ 
240936.717 (0.004) & 240936.742 (0.007) & -0.025 & 0.309 (0.002) &  1.22 (0.01) & CH$_3$OH & $5_{-2}-4_{-2} \, {\rm E} \, v_t = 2$\\ 
240938.950 (0.002) & 240938.974 (0.009) & -0.024 & 0.309 (0.002) &  2.31 (0.01) & CH$_3$OH & $5_{0}-4_{0} \, {\rm A^+} \, v_t = 2$\\ 
240948.316 (0.005) & 240948.343 (0.006) & -0.027 & 0.316 (0.007) &  1.01 (0.02) & CH$_3$OH & $5_{3}-4_{3} \, {\rm E} \, v_t = 2$\\ 
240952.043 (0.003) & 240952.056 (0.006) & -0.013 & 0.301 (0.004) &  1.50 (0.02) & CH$_3$OH & $5_{2}-4_{2} \, {\rm E} \, v_t = 2$\\ 
240958.897 (0.003) & 240958.922 (0.008) & -0.025 & 0.313 (0.002) &  2.02 (0.02) & CH$_3$OH & $5_{-1}-4_{-1} \, {\rm E} \, v_t = 2$\\ 
240960.531 (0.002) & 240960.557 (0.005) & -0.026 & 0.313 (0.002) &  3.91 (0.02) & CH$_3$OH & $5_{1}-4_{1} \, {\rm A^+} \, v_t = 1$\\ 
241042.690 (0.003) & 241042.589 (0.017) &  0.101 & 0.321 (0.004) &  1.53 (0.02) & CH$_3$OH & $22_{-6}-23_{-5} \, {\rm E} \, v_t = 0$\\ 
241159.170 (0.005) & 241159.199 (0.004) & -0.029 & 0.299 (0.006) &  1.30 (0.02) & CH$_3$OH & $5_{4}-4_{4} \, {\rm E} \, v_t = 1$\\ 
241166.555 (0.004) & 241166.580 (0.004) & -0.025 & 0.318 (0.005) &  1.87 (0.02) & CH$_3$OH & $5_{3}-4_{3} \, {\rm E} \, v_t = 1$\\ 
241178.435 (0.004) $^{\rm g}$ & 241178.445 (0.004) & -0.010 & 0.323 (0.003) &  1.70 (0.02) & CH$_3$OH & $5_{4}-4_{4} \, {\rm A^+} \, v_t = 1$\\ 
241178.435 (0.004) $^{\rm g}$ & 241178.445 (0.004) & -0.010 & 0.323 (0.003) &  1.70 (0.02) & CH$_3$OH & $5_{4}-4_{4} \, {\rm A^-} \, v_t = 1$\\ 
241179.862 (0.003) & 241179.886 (0.005) & -0.024 & 0.323 (0.003) &  2.54 (0.02) & CH$_3$OH & $5_{-3}-4_{-3} \, {\rm E} \, v_t = 1$\\ 
241184.182 (0.005) & 241184.189 (0.004) & -0.007 & 0.300 (0.006) &  1.13 (0.02) & CH$_3$OH & $5_{-4}-4_{-4} \, {\rm E} \, v_t = 1$\\ 
241187.410 (0.002) & 241187.428 (0.004) & -0.018 & 0.318 (0.003) &  2.82 (0.02) & CH$_3$OH & $5_{-2}-4_{-2} \, {\rm E} \, v_t = 1$\\ 
241192.837 (0.003) & 241192.856 (0.005) & -0.019 & 0.314 (0.004) &  3.53 (0.03) & CH$_3$OH & $5_{2}-4_{2} \, {\rm A^+} \, v_t = 1$\\ 
241196.406 (0.003) & 241196.430 (0.005) & -0.024 & 0.311 (0.003) &  3.48 (0.03) & CH$_3$OH & $5_{2}-4_{2} \, {\rm A^-} \, v_t = 1$\\ 
241198.281 (0.003) $^{\rm g}$ & 241198.285 (0.004) & -0.004 & 0.311 (0.003) &  3.99 (0.03) & CH$_3$OH & $5_{3}-4_{3} \, {\rm A^+} \, v_t = 1$\\ 
241198.281 (0.003) $^{\rm g}$ & 241198.291 (0.004) & -0.010 & 0.311 (0.003) &  3.99 (0.03) & CH$_3$OH & $5_{3}-4_{3} \, {\rm A^-} \, v_t = 1$\\ 
241203.683 (0.002) & 241203.706 (0.005) & -0.023 & 0.312 (0.002) &  4.12 (0.02) & CH$_3$OH & $5_{1}-4_{1} \, {\rm E} \, v_t = 1$\\ 
241206.008 (0.002) & 241206.035 (0.005) & -0.027 & 0.312 (0.002) &  4.16 (0.02) & CH$_3$OH & $5_{0}-4_{0} \, {\rm E} \, v_t = 1$\\ 
241210.754 (0.002) & 241210.764 (0.004) & -0.010 & 0.315 (0.002) &  2.58 (0.02) & CH$_3$OH & $5_{2}-4_{2} \, {\rm E} \, v_t = 1$\\ 
241238.133 (0.002) & 241238.144 (0.005) & -0.011 & 0.310 (0.003) &  2.77 (0.02) & CH$_3$OH & $5_{-1}-4_{-1} \, {\rm E} \, v_t = 1$\\ 
241267.847 (0.002) & 241267.862 (0.006) & -0.015 & 0.309 (0.003) &  2.69 (0.02) & CH$_3$OH & $5_{0}-4_{0} \, {\rm A^+} \, v_t = 1$\\ 
241281.695 (0.005) & 241283.133 (0.098) & -1.438 & 0.320 (0.007) &  0.89 (0.02) & CH$_3$OH & $34_{3}-34_{2} \, {\rm A^{-+}} \, v_t = 0$\\ 
241364.100 (0.004) & 241364.143 (0.011) & -0.043 & 0.310 (0.005) &  1.13 (0.01) & CH$_3$OH & $5_{1}-4_{1} \, {\rm A^-} \, v_t = 2$\\ 
241441.251 (0.002) & 241441.270 (0.005) & -0.019 & 0.312 (0.002) &  3.91 (0.03) & CH$_3$OH & $5_{1}-4_{1} \, {\rm A^-} \, v_t = 1$\\ 
241588.757 (0.002) & 241588.758 (0.022) & -0.001 & 0.318 (0.002) &  5.07 (0.03) & CH$_3$OH & $25_{3}-25_{2} \, {\rm A^{-+}} \, v_t = 0$\\ 
241700.174 (0.001) & 241700.159 (0.004) &  0.015 & 0.320 (0.002) &  9.93 (0.05) & CH$_3$OH & $5_{0}-4_{0} \, {\rm E} \, v_t = 0$\\ 
241767.250 (0.002) & 241767.234 (0.004) &  0.016 & 0.323 (0.002) &  9.84 (0.06) & CH$_3$OH & $5_{-1}-4_{-1} \, {\rm E} \, v_t = 0$\\ 
241791.373 (0.002) & 241791.352 (0.004) &  0.021 & 0.321 (0.002) & 10.58 (0.06) & CH$_3$OH & $5_{0}-4_{0} \, {\rm A^+} \, v_t = 0$\\ 
241806.522 (0.001) $^{\rm g}$ & 241806.524 (0.004) & -0.003 & 0.319 (0.002) &  6.05 (0.03) & CH$_3$OH & $5_{4}-4_{4} \, {\rm A^-} \, v_t = 0$\\ 
241806.522 (0.001) $^{\rm g}$ & 241806.525 (0.004) & -0.003 & 0.319 (0.002) &  6.05 (0.03) & CH$_3$OH & $5_{4}-4_{4} \, {\rm A^+} \, v_t = 0$\\ 
241813.263 (0.002) & 241813.255 (0.004) &  0.008 & 0.321 (0.003) &  2.95 (0.02) & CH$_3$OH & $5_{-4}-4_{-4} \, {\rm E} \, v_t = 0$\\ 
241829.637 (0.003) & 241829.629 (0.004) &  0.008 & 0.313 (0.003) &  2.93 (0.03) & CH$_3$OH & $5_{4}-4_{4} \, {\rm E} \, v_t = 0$\\ 
241832.928 (0.002) $^{\rm g}$ & 241832.718 (0.004) &  0.210 & 0.429 (0.002) &  9.13 (0.04) & CH$_3$OH & $5_{3}-4_{3} \, {\rm A^+} \, v_t = 0$\\ 
241832.928 (0.002) $^{\rm g}$ & 241833.106 (0.004) & -0.178 & 0.430 (0.002) &  9.13 (0.04) & CH$_3$OH & $5_{3}-4_{3} \, {\rm A^-} \, v_t = 0$\\ 
241842.302 (0.002) & 241842.284 (0.004) &  0.018 & 0.329 (0.002) &  8.23 (0.04) & CH$_3$OH & $5_{2}-4_{2} \, {\rm A^-} \, v_t = 0$\\ 
241843.605 (0.002) & 241843.604 (0.004) &  0.001 & 0.329 (0.002) &  6.15 (0.04) & CH$_3$OH & $5_{3}-4_{3} \, {\rm E} \, v_t = 0$\\ 
241852.309 (0.002) & 241852.299 (0.004) &  0.010 & 0.322 (0.002) &  5.84 (0.03) & CH$_3$OH & $5_{-3}-4_{-3} \, {\rm E} \, v_t = 0$\\ 
241879.044 (0.002) & 241879.025 (0.004) &  0.019 & 0.322 (0.002) & 10.45 (0.05) & CH$_3$OH & $5_{1}-4_{1} \, {\rm E} \, v_t = 0$\\ 
241887.687 (0.002) & 241887.674 (0.004) &  0.013 & 0.323 (0.002) &  8.49 (0.05) & CH$_3$OH & $5_{2}-4_{2} \, {\rm A^+} \, v_t = 0$\\ 
241904.151 (0.004) & 241904.147 (0.004) &  0.004 & 0.304 (0.000) &  9.60 (0.10) & CH$_3$OH & $5_{-2}-4_{-2} \, {\rm E} \, v_t = 0$\\ 
241904.668 (0.004) & 241904.643 (0.004) &  0.025 & 0.304 (0.000) &  9.52 (0.10) & CH$_3$OH & $5_{2}-4_{2} \, {\rm E} \, v_t = 0$\\ 
242286.305 (0.003) $^{\rm g}$ & 242286.264 (0.026) &  0.041 & 0.334 (0.004) &  1.20 (0.01) & CH$_3$OH & $27_{6}-26_{7} \, {\rm A^-} \, v_t = 0$\\ 
242286.305 (0.003) $^{\rm g}$ & 242286.419 (0.026) & -0.114 & 0.334 (0.004) &  1.20 (0.01) & CH$_3$OH & $27_{6}-26_{7} \, {\rm A^+} \, v_t = 0$\\ 
242446.171 (0.002) & 242446.084 (0.010) &  0.087 & 0.326 (0.002) &  4.91 (0.03) & CH$_3$OH & $14_{-1}-13_{-2} \, {\rm E} \, v_t = 0$\\ 
242490.276 (0.002) & 242490.245 (0.021) &  0.031 & 0.320 (0.002) &  5.39 (0.03) & CH$_3$OH & $24_{3}-24_{2} \, {\rm A^{-+}} \, v_t = 0$\\ 
243145.446 (0.004) & 243147.202 (0.119) & -1.756 & 0.309 (0.006) &  0.74 (0.01) & CH$_3$OH & $35_{3}-35_{2} \, {\rm A^{-+}} \, v_t = 0$\\ 
243397.548 (0.001) $^{\rm g}$ & 243397.393 (0.019) &  0.155 & 0.363 (0.002) &  3.26 (0.01) & CH$_3$OH & $18_{6}-19_{5} \, {\rm A^-} \, v_t = 0$\\ 
243397.548 (0.001) $^{\rm g}$ & 243397.654 (0.019) & -0.106 & 0.363 (0.002) &  3.26 (0.01) & CH$_3$OH & $18_{6}-19_{5} \, {\rm A^+} \, v_t = 0$\\ 
243412.658 (0.002) & 243412.610 (0.019) &  0.048 & 0.321 (0.002) &  6.68 (0.04) & CH$_3$OH & $23_{3}-23_{2} \, {\rm A^{-+}} \, v_t = 0$\\ 
243915.809 (0.002) & 243915.788 (0.004) &  0.021 & 0.321 (0.002) & 10.27 (0.05) & CH$_3$OH & $5_{1}-4_{1} \, {\rm A^-} \, v_t = 0$\\ 
243980.360 (0.004) & 243980.164 (0.031) &  0.196 & 0.318 (0.006) &  0.59 (0.01) & CH$_3$OH & $29_{5}-28_{6} \, {\rm A^+} \, v_t = 0$\\ 
243997.526 (0.006) & 243997.335 (0.031) &  0.191 & 0.352 (0.008) &  0.59 (0.01) & CH$_3$OH & $29_{5}-28_{6} \, {\rm A^-} \, v_t = 0$\\ 
244087.722 (0.011) & 244087.766 (0.073) & -0.044 & 0.296 (0.014) &  0.29 (0.01) & CH$_3$OH & $33_{-5}-32_{-6} \, {\rm E} \, v_t = 0$\\ 
244330.429 (0.002) & 244330.372 (0.018) &  0.057 & 0.319 (0.002) &  6.90 (0.04) & CH$_3$OH & $22_{3}-22_{2} \, {\rm A^{-+}} \, v_t = 0$\\ 
244337.956 (0.002) & 244337.983 (0.009) & -0.027 & 0.319 (0.002) &  3.43 (0.02) & CH$_3$OH & $9_{1}-8_{0} \, {\rm E} \, v_t = 1$\\ 
244652.651 (0.033) &  &  & 0.322 (0.042) &  0.10 (0.01) & UI &  \\ 
244996.453 (0.034) &  &  & 0.429 (0.044) &  0.13 (0.01) & UI &  \\ 
245094.395 (0.005) & 245094.503 (0.027) & -0.108 & 0.318 (0.006) &  0.79 (0.01) & CH$_3$OH & $18_{-6}-17_{-7} \, {\rm E} \, v_t = 1$\\ 
245223.082 (0.002) & 245223.019 (0.017) &  0.063 & 0.320 (0.002) &  8.40 (0.04) & CH$_3$OH & $21_{3}-21_{2} \, {\rm A^{-+}} \, v_t = 0$\\ 
245630.677 (0.006) & 245632.795 (0.145) & -2.118 & 0.306 (0.007) &  0.65 (0.01) & CH$_3$OH & $36_{3}-36_{2} \, {\rm A^{-+}} \, v_t = 0$\\ 
245646.555 (0.026) &  &  & 0.339 (0.033) &  0.14 (0.01) & UI &  \\ 
246074.655 (0.002) & 246074.605 (0.016) &  0.049 & 0.322 (0.002) & 10.18 (0.05) & CH$_3$OH & $20_{3}-20_{2} \, {\rm A^{-+}} \, v_t = 0$\\ 
246192.052 (0.026) & 246191.950 (0.029) &  0.102 & 0.277 (0.023) &  0.15 (0.01) & CH$_3$OH & $24_{4}-23_{3} \, {\rm A^+} \, v_t = 2$\\ 
246193.398 (0.025) & 246193.329 (0.029) &  0.069 & 0.277 (0.023) &  0.15 (0.01) & CH$_3$OH & $24_{4}-23_{3} \, {\rm A^-} \, v_t = 2$\\ 
246873.347 (0.002) & 246873.301 (0.015) &  0.046 & 0.321 (0.002) & 12.07 (0.06) & CH$_3$OH & $19_{3}-19_{2} \, {\rm A^{-+}} \, v_t = 0$\\ 
247162.049 (0.002) & 247161.950 (0.015) &  0.098 & 0.323 (0.002) &  5.24 (0.03) & CH$_3$OH & $16_{2}-15_{3} \, {\rm E} \, v_t = 0$\\ 
247228.661 (0.003) & 247228.587 (0.013) &  0.074 & 0.319 (0.003) &  3.13 (0.03) & CH$_3$OH & $4_{2}-5_{1} \, {\rm A^+} \, v_t = 0$\\ 
247610.970 (0.002) & 247610.918 (0.014) &  0.052 & 0.322 (0.002) & 13.10 (0.08) & CH$_3$OH & $18_{3}-18_{2} \, {\rm A^{-+}} \, v_t = 0$\\ 
247807.025 (0.030) & 247806.862 (0.195) &  0.163 & 0.276 (0.038) &  0.12 (0.01) & CH$_3$OH & $34_{11}-35_{10} \, {\rm A^+} \, v_t = 0$\\ 
247807.025 (0.030) & 247806.867 (0.195) &  0.158 & 0.276 (0.038) &  0.12 (0.01) & CH$_3$OH & $34_{11}-35_{10} \, {\rm A^-} \, v_t = 0$\\ 
247840.048 (0.002) & 247840.050 (0.017) & -0.002 & 0.320 (0.002) &  5.19 (0.03) & CH$_3$OH & $12_{-2}-13_{-3} \, {\rm E} \, v_t = 1$\\ 
247968.071 (0.002) & 247968.119 (0.024) & -0.048 & 0.319 (0.002) &  4.34 (0.03) & CH$_3$OH & $23_{1}-23_{0} \, {\rm E} \, v_t = 0$\\ 
247993.427 (0.030) & 247993.783 (0.429) & -0.356 & 0.236 (0.037) &  0.11 (0.01) & $^{13}$CH$_3$OH & $22_{3}-22_{2} \, {\rm A^{-+}} \, v_t = 0$\\ 
248001.727 (0.020) & 248001.871 (0.043) & -0.144 & 0.272 (0.025) &  0.19 (0.01) & CH$_3$OH & $23_{5}-24_{4} \, {\rm E} \, v_t = 1$\\ 
248282.445 (0.002) & 248282.424 (0.013) &  0.021 & 0.322 (0.002) & 13.30 (0.07) & CH$_3$OH & $17_{3}-17_{2} \, {\rm A^{-+}} \, v_t = 0$\\ 
248805.479 (0.009) & 248807.950 (0.176) & -2.471 & 0.295 (0.012) &  0.58 (0.02) & CH$_3$OH & $37_{3}-37_{2} \, {\rm A^{-+}} \, v_t = 0$\\ 
248854.922 (0.008) & 248854.996 (0.027) & -0.074 & 0.313 (0.010) &  0.70 (0.02) & CH$_3$OH & $15_{-1}-16_{-2} \, {\rm E} \, v_t = 1$\\ 
248885.481 (0.002) & 248885.468 (0.012) &  0.014 & 0.323 (0.002) & 14.20 (0.07) & CH$_3$OH & $16_{3}-16_{2} \, {\rm A^{-+}} \, v_t = 0$\\ 
248970.453 (0.004) & 248970.736 (0.143) & -0.283 & 0.322 (0.006) &  1.42 (0.02) & CH$_3$OH & $33_{-2}-33_{1} \, {\rm E} \, v_t = 0$\\ 
249003.848 (0.005) & 249004.019 (0.037) & -0.171 & 0.325 (0.006) &  1.29 (0.02) & CH$_3$OH & $10_{7}-9_{6} \, {\rm E} \, v_t = 1$\\ 
249192.810 (0.002) & 249192.836 (0.013) & -0.026 & 0.321 (0.003) &  4.40 (0.03) & CH$_3$OH & $16_{-3}-15_{-4} \, {\rm E} \, v_t = 0$\\ 
249419.934 (0.002) & 249419.924 (0.012) &  0.010 & 0.324 (0.002) & 14.40 (0.07) & CH$_3$OH & $15_{3}-15_{2} \, {\rm A^{-+}} \, v_t = 0$\\ 
249443.343 (0.003) & 249443.301 (0.014) &  0.042 & 0.325 (0.003) &  2.45 (0.02) & CH$_3$OH & $7_{4}-8_{3} \, {\rm A^-} \, v_t = 0$\\ 
249451.877 (0.003) & 249451.842 (0.014) &  0.035 & 0.325 (0.003) &  2.47 (0.02) & CH$_3$OH & $7_{4}-8_{3} \, {\rm A^+} \, v_t = 0$\\ 
249688.713 (0.058) & 249688.913 (0.284) & -0.200 & 0.352 (0.073) &  0.10 (0.02) & $^{13}$CH$_3$OH & $20_{3}-20_{2} \, {\rm A^{-+}} \, v_t = 0$\\ 
249887.477 (0.002) & 249887.467 (0.011) &  0.010 & 0.324 (0.002) & 15.00 (0.08) & CH$_3$OH & $14_{3}-14_{2} \, {\rm A^{-+}} \, v_t = 0$\\ 
250169.037 (0.024) & 250169.306 (0.047) & -0.269 & 0.336 (0.027) &  0.23 (0.02) & CH$_3$OH & $29_{-9}-30_{-8} \, {\rm E} \, v_t = 0$\\ 
250291.178 (0.002) & 250291.181 (0.010) & -0.003 & 0.324 (0.002) & 16.37 (0.08) & CH$_3$OH & $13_{3}-13_{2} \, {\rm A^{-+}} \, v_t = 0$\\ 
250455.175 (0.043) & 250455.289 (0.230) & -0.114 & 0.269 (0.053) &  0.12 (0.02) & $^{13}$CH$_3$OH & $19_{3}-19_{2} \, {\rm A^{-+}} \, v_t = 0$\\ 
250506.972 (0.002) & 250506.853 (0.011) &  0.119 & 0.325 (0.002) & 18.64 (0.09) & CH$_3$OH & $11_{0}-10_{1} \, {\rm A^+} \, v_t = 0$\\ 
250635.190 (0.002) & 250635.200 (0.010) & -0.010 & 0.325 (0.002) & 15.83 (0.08) & CH$_3$OH & $12_{3}-12_{2} \, {\rm A^{-+}} \, v_t = 0$\\ 
250924.391 (0.002) & 250924.398 (0.010) & -0.007 & 0.324 (0.002) & 17.29 (0.10) & CH$_3$OH & $11_{3}-11_{2} \, {\rm A^{-+}} \, v_t = 0$\\ 
250969.973 (0.002) & 250970.042 (0.029) & -0.069 & 0.320 (0.002) &  3.73 (0.02) & CH$_3$OH & $17_{3}-18_{4} \, {\rm E} \, v_t = 1$\\ 
251158.619 (0.022) & 251158.747 (0.186) & -0.128 & 0.229 (0.027) &  0.16 (0.02) & $^{13}$CH$_3$OH & $18_{3}-18_{2} \, {\rm A^{-+}} \, v_t = 0$\\ 
251164.091 (0.002) & 251164.108 (0.010) & -0.017 & 0.325 (0.002) & 15.80 (0.08) & CH$_3$OH & $10_{3}-10_{2} \, {\rm A^{-+}} \, v_t = 0$\\ 
251359.873 (0.002) & 251359.888 (0.011) & -0.015 & 0.324 (0.002) & 15.65 (0.08) & CH$_3$OH & $9_{3}-9_{2} \, {\rm A^{-+}} \, v_t = 0$\\ 
251517.290 (0.002) & 251517.309 (0.011) & -0.019 & 0.324 (0.002) & 14.54 (0.07) & CH$_3$OH & $8_{3}-8_{2} \, {\rm A^{-+}} \, v_t = 0$\\ 
251641.776 (0.002) & 251641.787 (0.012) & -0.011 & 0.324 (0.002) & 13.64 (0.07) & CH$_3$OH & $7_{3}-7_{2} \, {\rm A^{-+}} \, v_t = 0$\\ 
251709.851 (0.029) & 251709.789 (0.030) &  0.062 & 0.365 (0.036) &  0.14 (0.01) & CH$_3$OH & $9_{3}-10_{2} \, {\rm E} \, v_t = 2$\\ 
251738.420 (0.002) & 251738.437 (0.013) & -0.017 & 0.324 (0.002) & 11.76 (0.06) & CH$_3$OH & $6_{3}-6_{2} \, {\rm A^{-+}} \, v_t = 0$\\ 
251795.994 (0.023) & 251796.079 (0.149) & -0.085 & 0.317 (0.029) &  0.16 (0.01) & $^{13}$CH$_3$OH & $17_{3}-17_{2} \, {\rm A^{-+}} \, v_t = 0$\\ 
251811.946 (0.002) & 251811.956 (0.013) & -0.010 & 0.325 (0.002) &  9.58 (0.05) & CH$_3$OH & $5_{3}-5_{2} \, {\rm A^{-+}} \, v_t = 0$\\ 
251866.509 (0.002) & 251866.524 (0.014) & -0.015 & 0.326 (0.002) &  6.94 (0.04) & CH$_3$OH & $4_{3}-4_{2} \, {\rm A^{-+}} \, v_t = 0$\\ 
251890.874 (0.002) & 251890.886 (0.013) & -0.012 & 0.325 (0.002) &  9.65 (0.06) & CH$_3$OH & $5_{3}-5_{2} \, {\rm A^{+-}} \, v_t = 0$\\ 
251895.713 (0.001) & 251895.728 (0.013) & -0.015 & 0.322 (0.002) & 11.79 (0.06) & CH$_3$OH & $6_{3}-6_{2} \, {\rm A^{+-}} \, v_t = 0$\\ 
251900.440 (0.002) & 251900.452 (0.014) & -0.012 & 0.326 (0.002) &  7.16 (0.04) & CH$_3$OH & $4_{3}-4_{2} \, {\rm A^{+-}} \, v_t = 0$\\ 
251905.717 (0.002) & 251905.729 (0.014) & -0.012 & 0.326 (0.003) &  4.08 (0.03) & CH$_3$OH & $3_{3}-3_{2} \, {\rm A^{-+}} \, v_t = 0$\\ 
251917.051 (0.002) & 251917.065 (0.014) & -0.014 & 0.326 (0.003) &  4.00 (0.03) & CH$_3$OH & $3_{3}-3_{2} \, {\rm A^{+-}} \, v_t = 0$\\ 
251923.679 (0.002) & 251923.701 (0.012) & -0.022 & 0.324 (0.002) & 13.02 (0.07) & CH$_3$OH & $7_{3}-7_{2} \, {\rm A^{+-}} \, v_t = 0$\\ 
251984.819 (0.002) & 251984.837 (0.011) & -0.018 & 0.323 (0.002) & 14.30 (0.07) & CH$_3$OH & $8_{3}-8_{2} \, {\rm A^{+-}} \, v_t = 0$\\ 
252090.398 (0.002) & 252090.409 (0.011) & -0.011 & 0.325 (0.002) & 15.02 (0.07) & CH$_3$OH & $9_{3}-9_{2} \, {\rm A^{+-}} \, v_t = 0$\\ 
252106.788 (0.031) &  &  & 0.351 (0.038) &  0.10 (0.01) & UI &  \\ 
252252.840 (0.002) & 252252.849 (0.010) & -0.009 & 0.325 (0.002) & 16.16 (0.08) & CH$_3$OH & $10_{3}-10_{2} \, {\rm A^{+-}} \, v_t = 0$\\ 
252366.287 (0.013) & 252366.295 (0.120) & -0.008 & 0.339 (0.016) &  0.21 (0.01) & $^{13}$CH$_3$OH & $16_{3}-16_{2} \, {\rm A^{-+}} \, v_t = 0$\\ 
252485.672 (0.002) & 252485.675 (0.010) & -0.003 & 0.325 (0.002) & 16.26 (0.08) & CH$_3$OH & $11_{3}-11_{2} \, {\rm A^{+-}} \, v_t = 0$\\ 
252735.425 (0.009) & 252738.297 (0.216) & -2.872 & 0.308 (0.012) &  0.38 (0.01) & CH$_3$OH & $38_{3}-38_{2} \, {\rm A^{-+}} \, v_t = 0$\\ 
252803.398 (0.002) & 252803.388 (0.010) &  0.010 & 0.323 (0.002) & 16.32 (0.08) & CH$_3$OH & $12_{3}-12_{2} \, {\rm A^{+-}} \, v_t = 0$\\ 
252870.170 (0.016) & 252870.225 (0.095) & -0.055 & 0.267 (0.020) &  0.17 (0.01) & $^{13}$CH$_3$OH & $15_{3}-15_{2} \, {\rm A^{-+}} \, v_t = 0$\\ 
252974.057 (0.044) &  &  & 0.403 (0.055) &  0.08 (0.01) & UI &  \\ 
253078.988 (0.017) &  &  & 0.245 (0.021) &  0.15 (0.01) & UI &  \\ 
253221.390 (0.002) & 253221.376 (0.010) &  0.014 & 0.324 (0.002) & 15.48 (0.08) & CH$_3$OH & $13_{3}-13_{2} \, {\rm A^{+-}} \, v_t = 0$\\ 
253310.118 (0.021) & 253310.162 (0.076) & -0.044 & 0.279 (0.026) &  0.15 (0.01) & $^{13}$CH$_3$OH & $14_{3}-14_{2} \, {\rm A^{-+}} \, v_t = 0$\\ 
253689.524 (0.016) & 253689.530 (0.060) & -0.006 & 0.308 (0.019) &  0.18 (0.01) & $^{13}$CH$_3$OH & $13_{3}-13_{2} \, {\rm A^{-+}} \, v_t = 0$\\ 
253755.840 (0.002) & 253755.809 (0.011) &  0.031 & 0.324 (0.002) & 14.52 (0.08) & CH$_3$OH & $14_{3}-14_{2} \, {\rm A^{+-}} \, v_t = 0$\\ 
254012.579 (0.028) & 254012.582 (0.047) & -0.003 & 0.259 (0.035) &  0.16 (0.02) & $^{13}$CH$_3$OH & $12_{3}-12_{2} \, {\rm A^{-+}} \, v_t = 0$\\ 
254015.414 (0.003) & 254015.377 (0.012) &  0.037 & 0.318 (0.004) &  1.41 (0.01) & CH$_3$OH & $2_{0}-1_{-1} \, {\rm E} \, v_t = 0$\\ 
254284.162 (0.036) & 254284.138 (0.037) &  0.024 & 0.243 (0.045) &  0.16 (0.02) & $^{13}$CH$_3$OH & $11_{3}-11_{2} \, {\rm A^{-+}} \, v_t = 0$\\ 
254419.348 (0.002) & 254419.419 (0.015) & -0.071 & 0.313 (0.003) &  2.38 (0.02) & CH$_3$OH & $11_{5}-12_{4} \, {\rm E} \, v_t = 0$\\ 
254423.568 (0.002) & 254423.520 (0.012) &  0.048 & 0.323 (0.002) & 13.19 (0.07) & CH$_3$OH & $15_{3}-15_{2} \, {\rm A^{+-}} \, v_t = 0$\\ 
254509.313 (0.017) & 254509.364 (0.050) & -0.051 & 0.283 (0.021) &  0.19 (0.01) & $^{13}$CH$_3$OH & $10_{3}-10_{2} \, {\rm A^{-+}} \, v_t = 0$\\ 
254693.548 (0.024) & 254693.481 (0.050) &  0.067 & 0.384 (0.030) &  0.14 (0.01) & $^{13}$CH$_3$OH & $9_{3}-9_{2} \, {\rm A^{-+}} \, v_t = 0$\\ 
254841.831 (0.035) & 254841.818 (0.050) &  0.013 & 0.420 (0.025) &  0.14 (0.01) & $^{13}$CH$_3$OH & $8_{3}-8_{2} \, {\rm A^{-+}} \, v_t = 0$\\ 
254959.472 (0.024) & 254959.398 (0.050) &  0.074 & 0.314 (0.030) &  0.15 (0.01) & $^{13}$CH$_3$OH & $7_{3}-7_{2} \, {\rm A^{-+}} \, v_t = 0$\\ 
255051.053 (0.027) & 255050.965 (0.050) &  0.088 & 0.274 (0.033) &  0.14 (0.01) & $^{13}$CH$_3$OH & $6_{3}-6_{2} \, {\rm A^{-+}} \, v_t = 0$\\ 
255173.080 (0.063) & 255173.019 (0.050) &  0.061 & 0.546 (0.082) &  0.06 (0.01) & $^{13}$CH$_3$OH & $4_{3}-4_{2} \, {\rm A^{-+}} \, v_t = 0$\\ 
255192.418 (0.030) & 255192.375 (0.050) &  0.043 & 0.341 (0.021) &  0.09 (0.01) & $^{13}$CH$_3$OH & $5_{3}-5_{2} \, {\rm A^{+-}} \, v_t = 0$\\ 
255193.508 (0.026) & 255193.509 (0.050) & -0.001 & 0.341 (0.021) &  0.10 (0.01) & $^{13}$CH$_3$OH & $6_{3}-6_{2} \, {\rm A^{+-}} \, v_t = 0$\\ 
255203.764 (0.026) & 255203.728 (0.050) &  0.035 & 0.271 (0.032) &  0.09 (0.01) & $^{13}$CH$_3$OH & $4_{3}-4_{2} \, {\rm A^{+-}} \, v_t = 0$\\ 
255214.894 (0.029) & 255214.891 (0.050) &  0.003 & 0.286 (0.036) &  0.11 (0.01) & $^{13}$CH$_3$OH & $7_{3}-7_{2} \, {\rm A^{+-}} \, v_t = 0$\\ 
255220.980 (0.049) & 255220.865 (0.050) &  0.115 & 0.275 (0.060) &  0.06 (0.01) & $^{13}$CH$_3$OH & $3_{3}-3_{2} \, {\rm A^{+-}} \, v_t = 0$\\ 
255241.954 (0.002) & 255241.888 (0.013) &  0.066 & 0.323 (0.002) & 11.61 (0.06) & CH$_3$OH & $16_{3}-16_{2} \, {\rm A^{+-}} \, v_t = 0$\\ 
255265.659 (0.020) & 255265.637 (0.050) &  0.022 & 0.330 (0.025) &  0.13 (0.01) & $^{13}$CH$_3$OH & $8_{3}-8_{2} \, {\rm A^{+-}} \, v_t = 0$\\ 
255355.914 (0.019) & 255355.930 (0.050) & -0.016 & 0.296 (0.024) &  0.16 (0.01) & $^{13}$CH$_3$OH & $9_{3}-9_{2} \, {\rm A^{+-}} \, v_t = 0$\\ 
255394.750 (0.009) &  &  & 0.334 (0.011) &  0.37 (0.01) & UI &  \\ 
255496.977 (0.018) & 255496.966 (0.050) &  0.011 & 0.327 (0.023) &  0.16 (0.01) & $^{13}$CH$_3$OH & $10_{3}-10_{2} \, {\rm A^{+-}} \, v_t = 0$\\ 
255700.998 (0.018) & 255701.008 (0.038) & -0.010 & 0.313 (0.022) &  0.15 (0.01) & $^{13}$CH$_3$OH & $11_{3}-11_{2} \, {\rm A^{+-}} \, v_t = 0$\\ 
255742.658 (0.022) & 255742.822 (0.063) & -0.164 & 0.307 (0.027) &  0.13 (0.01) & CH$_3$OH & $27_{-2}-26_{-1} \, {\rm E} \, v_t = 1$\\ 
255877.295 (0.050) & 255873.476 (0.501) &  3.819 & 0.303 (0.061) &  0.06 (0.01) & CH$_3$OH & $37_{3}-38_{2} \, {\rm A^-} \, v_t = 1$\\ 
255981.138 (0.015) & 255981.193 (0.050) & -0.055 & 0.284 (0.019) &  0.18 (0.01) & $^{13}$CH$_3$OH & $12_{3}-12_{2} \, {\rm A^{+-}} \, v_t = 0$\\ 
256228.802 (0.002) & 256228.714 (0.014) &  0.088 & 0.321 (0.002) & 12.50 (0.06) & CH$_3$OH & $17_{3}-17_{2} \, {\rm A^{+-}} \, v_t = 0$\\ 
256351.455 (0.027) & 256351.482 (0.064) & -0.027 & 0.358 (0.034) &  0.15 (0.01) & $^{13}$CH$_3$OH & $13_{3}-13_{2} \, {\rm A^{+-}} \, v_t = 0$\\ 
256826.474 (0.032) & 256826.572 (0.083) & -0.098 & 0.302 (0.040) &  0.17 (0.02) & $^{13}$CH$_3$OH & $14_{3}-14_{2} \, {\rm A^{+-}} \, v_t = 0$\\ 
257402.187 (0.002) & 257402.086 (0.015) &  0.101 & 0.319 (0.002) & 11.41 (0.05) & CH$_3$OH & $18_{3}-18_{2} \, {\rm A^{+-}} \, v_t = 0$\\ 
257421.703 (0.030) & 257421.792 (0.106) & -0.089 & 0.305 (0.036) &  0.16 (0.02) & $^{13}$CH$_3$OH & $15_{3}-15_{2} \, {\rm A^{+-}} \, v_t = 0$\\ 
257482.704 (0.013) & 257485.943 (0.264) & -3.239 & 0.274 (0.016) &  0.30 (0.01) & CH$_3$OH & $39_{3}-39_{2} \, {\rm A^{-+}} \, v_t = 0$\\ 
257518.311 (0.019) & 257518.019 (0.081) &  0.292 & 0.323 (0.024) &  0.20 (0.01) & CH$_3$OH & $29_{-6}-28_{-5} \, {\rm E} \, v_t = 1$\\ 
258152.898 (0.028) & 258153.004 (0.135) & -0.106 & 0.312 (0.034) &  0.16 (0.01) & $^{13}$CH$_3$OH & $16_{3}-16_{2} \, {\rm A^{+-}} \, v_t = 0$\\ 
258780.385 (0.002) & 258780.248 (0.016) &  0.137 & 0.320 (0.002) & 11.07 (0.05) & CH$_3$OH & $19_{3}-19_{2} \, {\rm A^{+-}} \, v_t = 0$\\ 
259036.348 (0.026) & 259036.489 (0.170) & -0.141 & 0.337 (0.031) &  0.14 (0.01) & $^{13}$CH$_3$OH & $17_{3}-17_{2} \, {\rm A^{+-}} \, v_t = 0$\\ 
259273.622 (0.002) & 259273.686 (0.032) & -0.064 & 0.314 (0.002) &  3.51 (0.02) & CH$_3$OH & $17_{2}-16_{1} \, {\rm A^-} \, v_t = 1$\\ 
259581.341 (0.002) & 259581.398 (0.024) & -0.057 & 0.315 (0.002) &  3.86 (0.02) & CH$_3$OH & $24_{1}-24_{0} \, {\rm E} \, v_t = 0$\\ 
260064.206 (0.004) & 260064.318 (0.023) & -0.112 & 0.308 (0.005) &  1.02 (0.01) & CH$_3$OH & $20_{-8}-21_{-7} \, {\rm E} \, v_t = 0$\\ 
260088.600 (0.024) & 260088.839 (0.214) & -0.239 & 0.300 (0.029) &  0.14 (0.01) & $^{13}$CH$_3$OH & $18_{3}-18_{2} \, {\rm A^{+-}} \, v_t = 0$\\ 
260381.620 (0.002) & 260381.463 (0.017) &  0.157 & 0.319 (0.002) & 10.57 (0.05) & CH$_3$OH & $20_{3}-20_{2} \, {\rm A^{+-}} \, v_t = 0$\\ 
261061.344 (0.002) & 261061.320 (0.015) &  0.024 & 0.317 (0.003) &  2.75 (0.02) & CH$_3$OH & $21_{-4}-20_{-5} \, {\rm E} \, v_t = 0$\\ 
261124.451 (0.026) &  &  & 0.260 (0.031) &  0.15 (0.01) & UI &  \\ 
261326.553 (0.026) & 261326.838 (0.267) & -0.285 & 0.279 (0.031) &  0.13 (0.01) & $^{13}$CH$_3$OH & $19_{3}-19_{2} \, {\rm A^{+-}} \, v_t = 0$\\ 
261704.451 (0.003) & 261704.409 (0.015) &  0.042 & 0.315 (0.003) &  2.33 (0.02) & CH$_3$OH & $12_{6}-13_{5} \, {\rm E} \, v_t = 0$\\ 
261805.710 (0.002) & 261805.675 (0.006) &  0.035 & 0.319 (0.002) &  4.12 (0.02) & CH$_3$OH & $2_{1}-1_{0} \, {\rm E} \, v_t = 0$\\ 
262224.062 (0.002) & 262223.872 (0.018) &  0.190 & 0.318 (0.002) &  9.54 (0.05) & CH$_3$OH & $21_{3}-21_{2} \, {\rm A^{+-}} \, v_t = 0$\\ 
262767.059 (0.036) & 262767.346 (0.332) & -0.287 & 0.319 (0.043) &  0.12 (0.01) & $^{13}$CH$_3$OH & $20_{3}-20_{2} \, {\rm A^{+-}} \, v_t = 0$\\ 
262998.943 (0.012) & 262998.917 (0.034) &  0.026 & 0.331 (0.014) &  0.65 (0.02) & CH$_3$OH & $30_{4}-29_{5} \, {\rm A^-} \, v_t = 0$\\ 
263104.903 (0.029) & 263108.495 (0.322) & -3.592 & 0.300 (0.035) &  0.23 (0.02) & CH$_3$OH & $40_{3}-40_{2} \, {\rm A^{-+}} \, v_t = 0$\\ 
263793.881 (0.001) & 263793.875 (0.017) &  0.006 & 0.315 (0.002) &  4.43 (0.02) & CH$_3$OH & $5_{1}-6_{2} \, {\rm A^+} \, v_t = 1$\\ 
263815.883 (0.011) &  &  & 0.338 (0.013) &  0.40 (0.01) & UI &  \\ 
\end{longtable}
}

\end{document}